\documentclass[fleqn,usenatbib]{mnras}

\usepackage{newtxtext,newtxmath}

\usepackage[T1]{fontenc}

\DeclareRobustCommand{\VAN}[3]{#2}
\let\VANthebibliography\thebibliography
\def\thebibliography{\DeclareRobustCommand{\VAN}[3]{##3}\VANthebibliography}

\usepackage{graphicx}	
\usepackage{amsmath}	

\title[Magnetic fields in PPNe with JCMT POL-2]{JCMT POL-2 observations of magnetic fields potentially shaped by outflows in the pre-planetary nebulae CRL 618 and OH231.8+4.2}

\author[K. Pattle \& G. Savini]{
Kate Pattle$^{1}$\thanks{E-mail: k.pattle@ucl.ac.uk (KP)}
and Giorgio Savini$^{1}$
\\
$^{1}$Department of Physics and Astronomy, University College London, Gower Street, London WC1E 6BT, United Kingdom
}

\date{Accepted 2025 July 29. Received 2025 June 25; in original form 2024 April 13}

\pubyear{2024}

\begin{document}
\label{firstpage}
\pagerange{\pageref{firstpage}--\pageref{lastpage}}
\maketitle

\begin{abstract}
We present the first observations of magnetic fields in {pre-planetary nebulae (PPNe)} made with the POL-2 polarimeter on the James Clerk Maxwell Telescope (JCMT).  We observed the PPNe CRL 618 and OH231.8+4.2 in 850\,$\mu$m polarized light.  In both cases, we observe ordered magnetic fields that appear to arise from dusty circumstellar material that has been swept up by the passage of outflows driven by the central post-Asymptotic Giant Branch (post-AGB) star.  CRL 618 shows a magnetic field aligned with one of the most extreme position angles of the outflowing bullets ejected from the central source. We hypothesize that polarized emission in CRL 618 may preferentially arise from material in the walls of the dust cavity opened by the ejected bullets.  Conversely, OH231.8+4.2 shows a magnetic field that is aligned approximately perpendicular to the outflow direction, which may preferentially arise from an infrared-bright dense clump embedded near the base of the outflow.  Despite CRL 618 being carbon-rich and OH231.8 being oxygen-rich, there is no significant difference in the polarization fractions of the two sources.  This suggests that at linear resolutions $\sim 10^{4}$\,au, {the complexity of the} magnetic field geometry {on scales smaller than the beam}, rather than grain composition, sets the measured polarization fraction of these sources.

\end{abstract}




\section{Introduction}

Pre-planetary nebulae (PPNe; also known as protoplanetary nebulae) are a brief ($\sim 10^3$\,yr; e.g. \citealt{balick2019}), far-infrared-bright, phase in the evolution of Asymptotic Giant Branch (AGB) stars into planetary nebulae (e.g. \citealt{kwok1993}).  PPNe consist of shells of molecular gas and dust, relics of the AGB star's mass loss phase, which are being shocked by collimated outflows from the star \citep[e.g.][]{ueta2000,sahai2007}.  These objects thus present an excellent opportunity to study key sites of dust production in the modern Universe.

CRL 618 (the Westbrook Nebula; driven by V353 Aur) and OH231.8+4.2 (hereafter OH231.8; the Calabash or Rotten Egg Nebula; driven by QX Pup) are two nearby (0.9 and 1.54 kpc; \citealt{goodrich1991,choi2012}), young (200 and 770 yr; \citealt{kwok1984,alcolea2001}) well-studied {PPNe}.  Both sources present an abundance of dust emission \citep[e.g.][]{knapp1993} and molecular lines \citep[e.g.][]{lee2013,sanchezcontreras2015} arising from their circumstellar envelope (CSE), as expected for PPNe as a result of mass-loss events during their AGB phase \citep[e.g.][]{parthasarathy1986}. While CRL 618 is rich in carbonaceous dust and molecules \citep[C-rich;][]{bujarrabal1988,chiar1998}, OH231.8 is more oxygen-rich in its chemistry \citep[O-rich;][]{morris1987} with a higher content of larger dust grain size \citep{sabin2014}.  However, both sources have $^{12}$CO outflows: CRL 618 has a fast CO outflow \citep{cernicharo1989} with several dynamically ejected bullets \citep{huang2016}, while OH231.8 has a well-collimated $^{12}$CO jet \citep{alcolea2001} and a bipolar outflow with shocks \citep{reipurth1987}, with complex wind structure on small scales \citep{sanchezcontreras2022}.

Both CRL 618 and OH231.8 are known to have magnetic fields in their CSEs: \citet{duthu2017} measured a line-of-sight (LOS) magnetic field strength of $\approx 6$\,mG for CRL 618 at a distance of 2700 AU from the central star using Institut de Radioastronomie Millim\'etrique (IRAM) 30m Telescope CN Zeeman measurements, while \citet{lealferreira2012} measured a LOS field strength of $\approx 45$\,mG at a few tens of AU from the central star of OH231.8 using National Radio Astronomy Observatory (NRAO) H$_2$O maser Zeeman observations.  Interferometric dust polarization observations of these two objects (\citealt{sabin2014}; \citealt{sabin2015}; \citealt{sabin2020}) suggest that the directions of their magnetic fields are correlated with the directions of their outflows.
However, these observations, made with the Submillimeter Array (SMA; CRL 618 and OH231.8; \citealt{sabin2014}), the Combined Array for Research in Millimeter-wave Astronomy (CARMA; OH231.8; \citealt{sabin2015}), and the Atacama Large Millimeter/submillimeter Array (ALMA; OH231.8; \citealt{sabin2020}), have been restricted to the central few arcseconds around the central stars, as are the Zeeman measurements described above.  Single-dish observations have been made of polarized emission from dust in both AGB star envelopes \citep{andersson2024} and in planetary nebulae \citep{greaves2002} using the James Clerk Maxwell Telescope (JCMT). 
Similar single-dish observations of the PPNe CRL 618 and OH231.8 may thus allow us to probe their magnetic fields, and the interaction of these magnetic fields with their outflows, on larger scales than are recovered in interferometric observations of these objects.

Observing the polarized submillimetre dust emission in the {PPNe} arising from C- and O-rich AGB stars -- such as CRL 618 and OH231.8, respectively -- may allow us to investigate the polarization properties of silicate and carbonaceous dust grains in their sites of formation.  C-rich stars are expected to preferentially produce carbonaceous dust (amorphous carbon and graphite), whilst O-rich stars preferentially form silicates (e.g. \citealt{ishihara2011}).  Dust grains are preferentially aligned with respect to the magnetic field, producing linearly polarized emission: silicate grains are paramagnetic, and so should be more easily aligned with the magnetic field, thereby producing a stronger polarization signal \citep{hoang2016}.  Resolved SMA observations found that CRL 618 is on average 0.3\% polarized, while OH231.8 is on average 4.3\% polarized at 850\,$\mu$m near their centres \citep{sabin2014}, supporting this hypothesis, but this effect has not yet been observed in the larger CSEs.

In this work, {we investigate} the magnetic field and dust grain properties of the {PPNe} CRL 618 and OH231.8 on $\sim 10^{4}-10^{5}$ au scales using 850\,$\mu$m polarized dust emission observations made using the POL-2 polarimeter on the JCMT.  Section~\ref{sec:obs} describes our observations and the data reduction process. Section~\ref{sec:results} describes our results. Section~\ref{sec:discussion}
compares our observations to previous measurements, and interprets our results. Our conclusions are presented in Section~\ref{sec:conclusion}.

\begin{figure*}
    \centering
    \includegraphics[width=\textwidth]{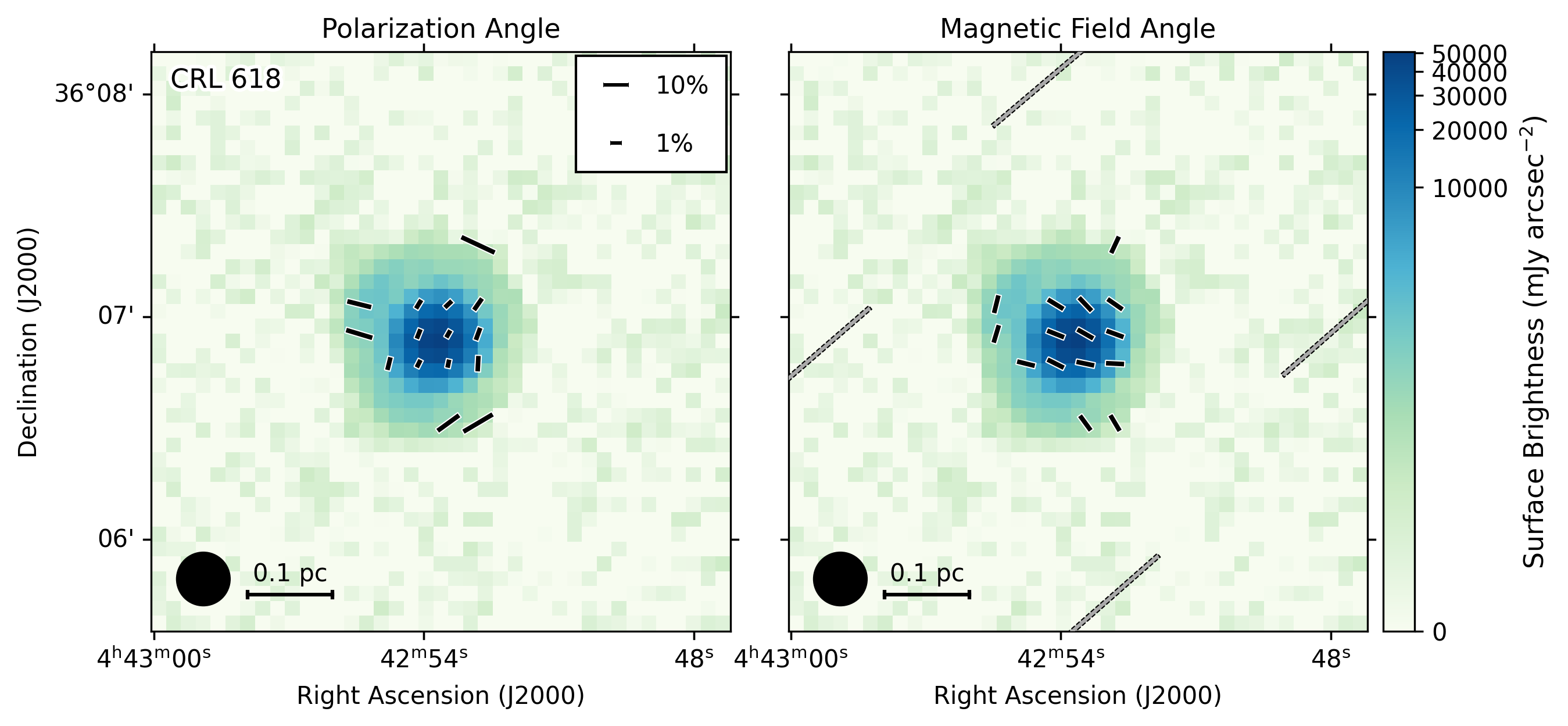}
    \caption{Our JCMT POL-2 observations of CRL 618.  Polarization and magnetic field vectors are overplotted on our POL-2 850\,$\mu$m Stokes $I$ (total intensity) image, the colour scale for which is shown to the right of the images. The Stokes $I$ images are shown on the default 4-arcsec pixel grid, with vectors overplotted on the 8-arcsec pixel grid used for analysis.  \textit{Left:} Black vectors with white outlines are POL-2 850 $\mu$m polarization vectors, with lengths proportional to their polarization fraction, as shown in the key in the upper right of the panel.  \textit{Right:} Black vectors with white outlines are POL-2 850\,$\mu$m magnetic field vectors (polarization vectors rotated by 90$^{\circ}$); grey vectors with dashed black outlines are \textit{Planck} 353\,GHz (850\,$\mu$m) magnetic field vectors, with 5$^{\prime}$ resolution.  Both JCMT and \textit{Planck} magnetic field vectors are plotted with constant length.}
    \label{fig:crl618}
\end{figure*}

\begin{figure*}
    \centering
    \includegraphics[width=\textwidth]{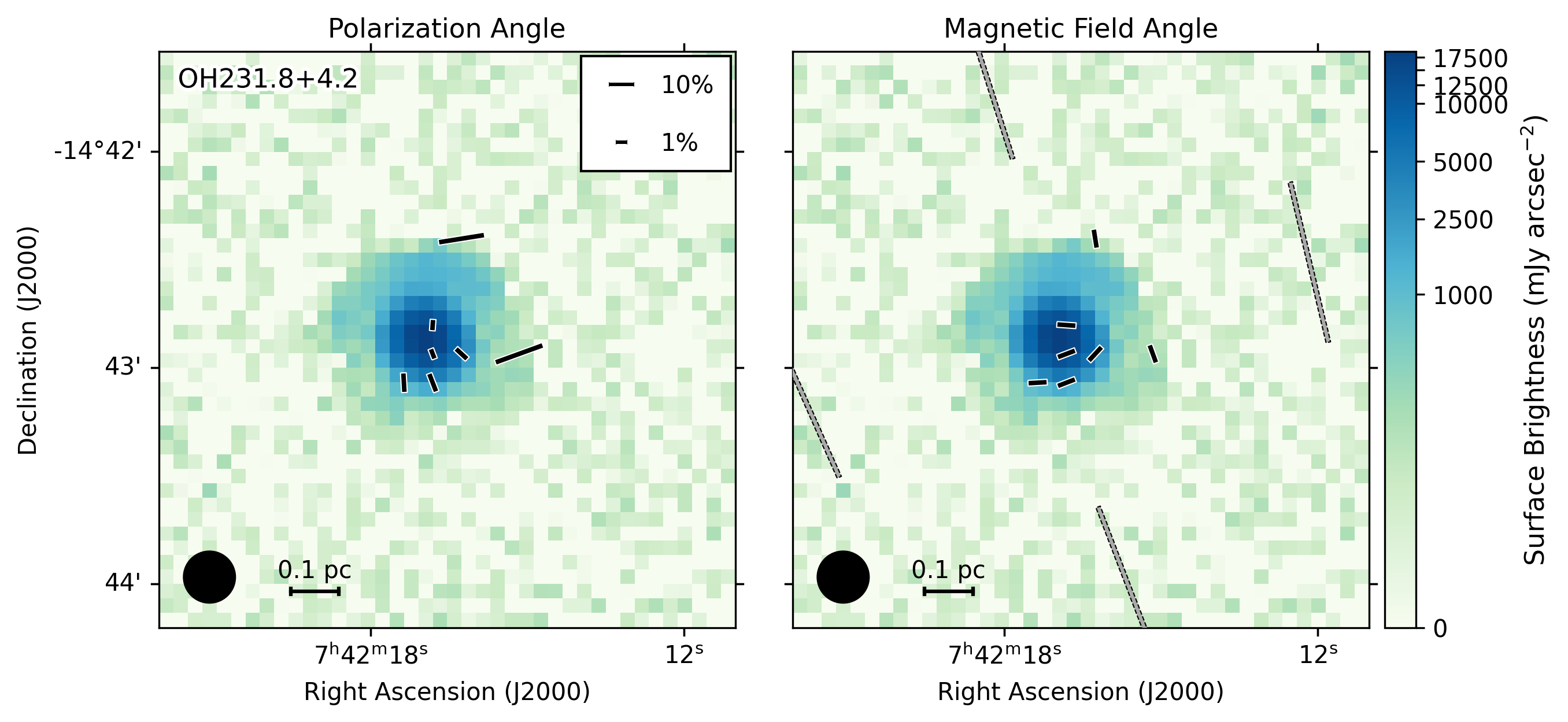}
    \caption{Our JCMT POL-2 observations of OH231.8.  Polarization and magnetic field vectors are overplotted on our POL-2 850\,$\mu$m Stokes $I$ (total intensity) image, the colour scale for which is shown to the right of the images. The Stokes $I$ images are shown on the default 4-arcsec pixel grid, with vectors overplotted on the 8-arcsec pixel grid used for analysis.   \textit{Left:} Black vectors with white outlines are POL-2 850 $\mu$m polarization vectors, with lengths proportional to their polarization fraction, , as shown in the key in the upper right of the panel.  \textit{Right:} Black vectors with white outlines are POL-2 850\,$\mu$m magnetic field vectors (polarization vectors rotated by 90$^{\circ}$); grey vectors with dashed black outlines are \textit{Planck} 353\,GHz (850\,$\mu$m) magnetic field vectors, with 5$^{\prime}$ resolution. Both JCMT and \textit{Planck} magnetic field vectors are plotted with constant length.  The Stokes $I$ images are shown on the default 4-arcsec pixel grid, with vectors overplotted on the 8-arcsec pixel grid used for analysis.}
    \label{fig:oh231}
\end{figure*}

\section{Observations}
\label{sec:obs}

\begin{figure}
    \centering
    \includegraphics[width=0.47\textwidth]{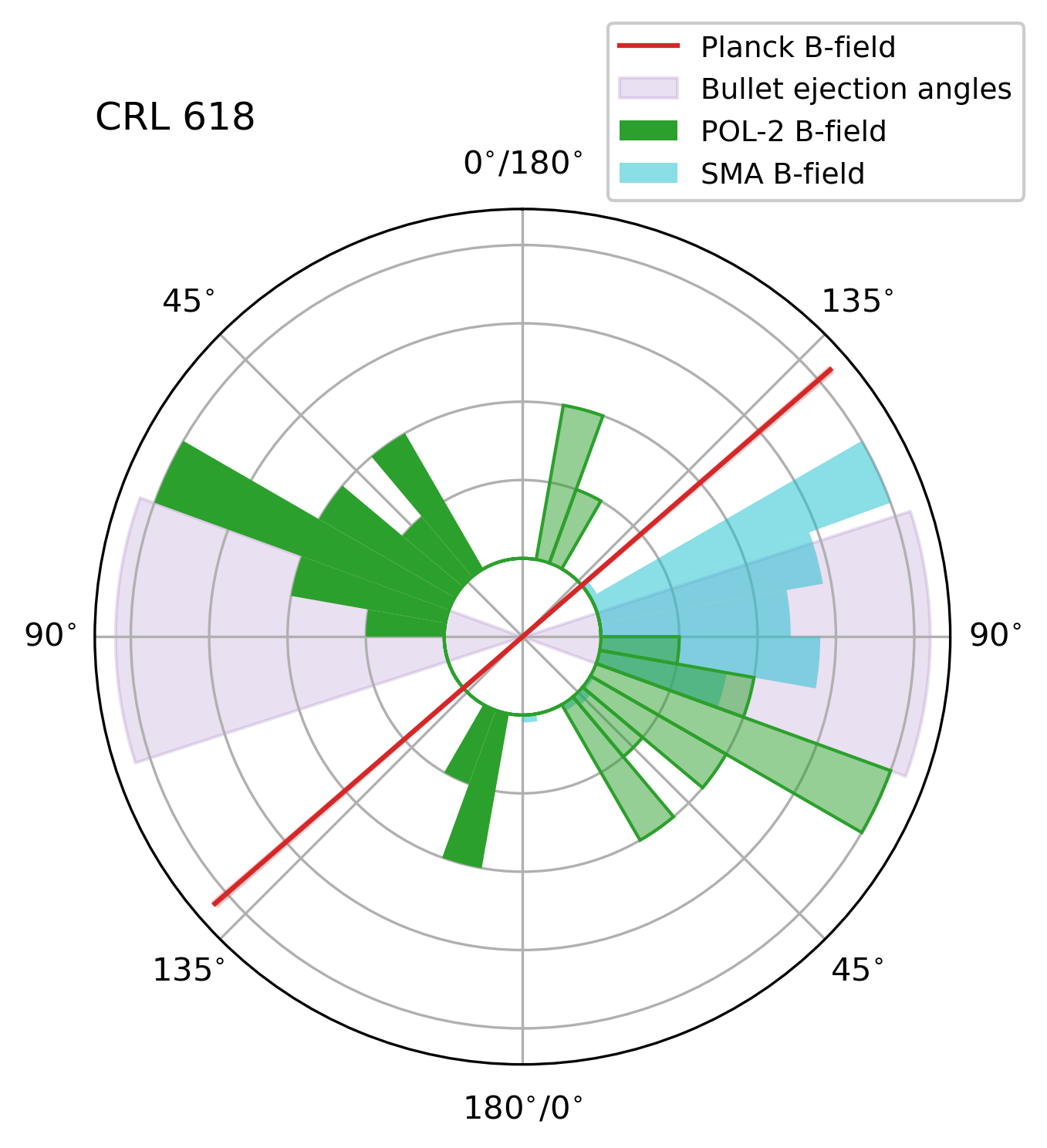}
    \caption{A polar histogram of magnetic field angle in CRL 618.  The POL-2 histogram is shown in green, while the histogram of magnetic field angles observed using the SMA \citep{sabin2014} is shown in cyan.  The mean \textit{Planck} magnetic field angle, {$130.9^{\circ}\pm0.4^{\circ}$ E of N} is shown as a red line, with its standard deviation shown as a shaded red sector (note that the standard deviation is sufficiently small that the red sector is obscured by the line showing the mean field direction).  The spread in position angle of the bullet ejections, $70-108^{\circ}$ E of N \citep{huang2016}, which we refer to as the `outflow cone angle', is shown as a light purple sector.  The left-hand side of the plot shows the POL-2 magnetic field histogram only.  The right-hand side of the plot overplots the POL-2 and the SMA data.  Due to the $\pm 180^{\circ}$ ambiguity on polarization vector measurements, opposite angles on the plot agree. In Cartesian space, the area of the POL-2 histogram is normalised to 1, while the SMA histogram is scaled so that its maximum value agrees with the maximum POL-2 histogram value.  The projection of the histograms on to a circle means that their areas are distorted.}
    \label{fig:histogram_crl618}
\end{figure}

We observed CRL 618 3 times consecutively on 2023 Aug 11, and OH231.8 3 times consecutively on 2023 Aug 31 using the POL-2 polarimeter \citep{friberg2016} mounted on the Submillimetre Common-User Bolometer Array 2 (SCUBA-2; \citealt{holland2013}) on the James Clerk Maxwell Telescope (JCMT).  The data were taken in JCMT Weather Band 1 ($\tau_{225\,{\rm GHz}}<0.05$), under project code M23BP031.  Each observation consisted of a POL-2 DAISY scan pattern, lasting 38 minutes each for CRL 618 and 32 minutes each for OH231.8+4.2.  

The data were reduced using the $pol2map$\footnote{\url{http://starlink.eao.hawaii.edu/docs/sun258.htx/sun258ss73.html}} script, from the \textsc{Smurf} package in the \textit{Starlink} software suite \citep{chapin2013}.  See \citet{pattle2021} for a detailed description of the POL-2 data reduction process.  The only non-standard parameter used in the data reduction process was setting `mapvar=no' in the final call to \textit{pol2map}, as is advised when reducing a small number of repeats of a target\footnote{\url{https://starlink.eao.hawaii.edu/docs/sc22.pdf}}.

Instrumental polarisation (IP) was corrected for using the `August 2019' IP model\footnote{\url{https://www.eaobservatory.org/jcmt/2019/08/new-ip-models-for-pol2-data/}}.  {The IP induced by the JCMT on POL-2 observations varies as a function of elevation.  Our two sources were observed over elevation ranges of 35--61$^{\circ}$ for CRL 618 and 48--55$^{\circ}$ for OH231.8, with very similar mean elevations of 48$^{\circ}$ for CRL 618 and 52$^{\circ}$ for OH231.8.  At these elevations, the Stokes $Q$ IP is in the range --(0.5--0.75)\%, while the Stokes $U$ IP is in the range --(2--2.25)\%.  This artificial polarization signal is removed from our Stokes $Q$ and $U$ maps by the \textit{pol2map} routine before any analysis is performed.}

The 850$\mu$m data were calibrated using a flux conversion factor (FCF) of 2795 mJy\,arcsec$^{-2}$\,pW$^{-1}$ using the post-2018 June 30 SCUBA-2 FCF of 2070 mJy\,arcsec$^{-2}$\,pW$^{-1}$ \citep{mairs2021} multiplied by a factor of 1.35 to account for additional losses in POL-2 \citep{friberg2016}.  POL-2 observes simultaneously at 850$\mu$m and 450$\mu$m, but we consider only the 850$\mu$m observations in this work.

\begin{figure}
    \centering
    \includegraphics[width=0.47\textwidth]{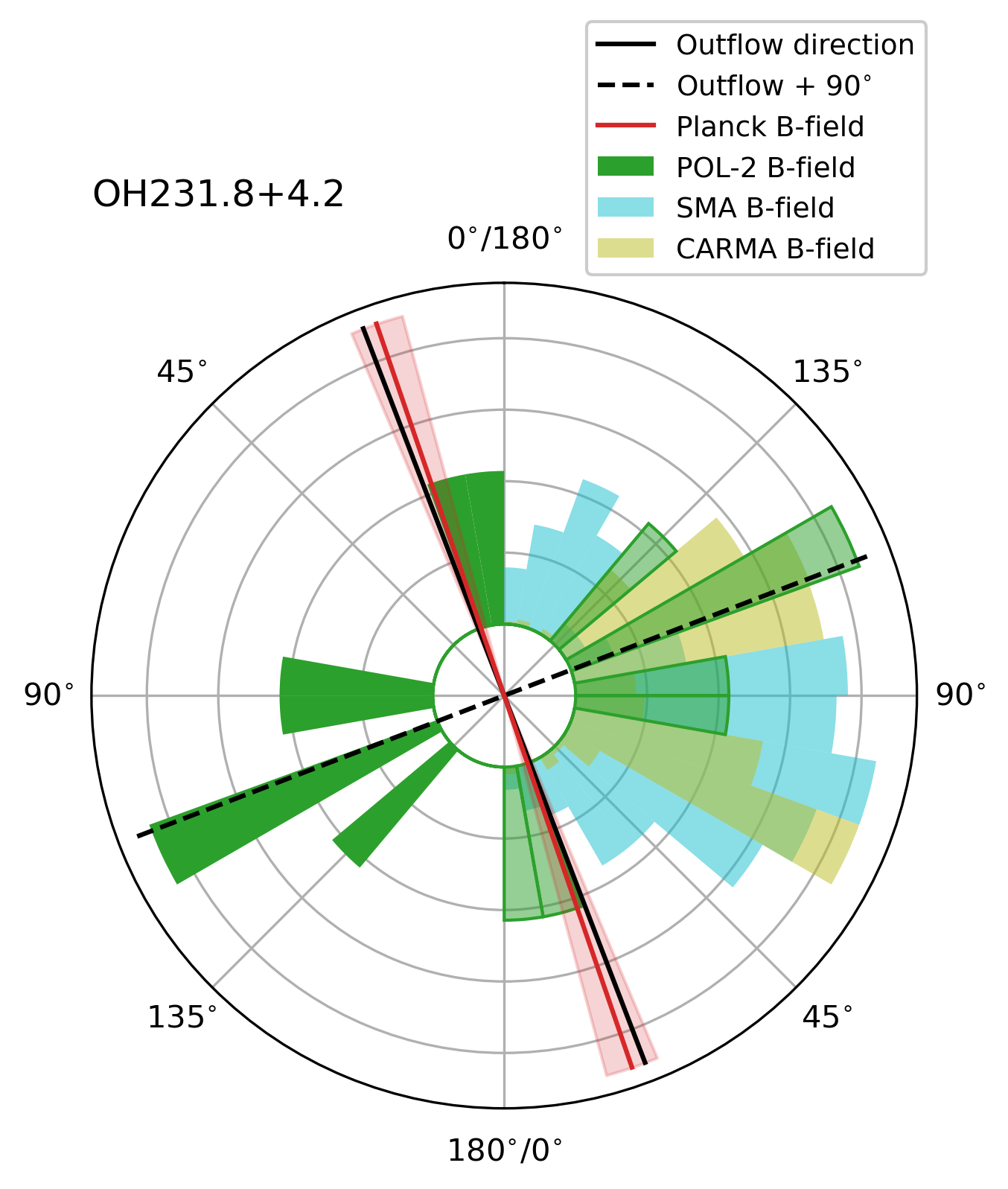}
    \caption{A polar histogram of magnetic field angle in OH231.8.  The POL-2 histogram is shown in green, while the histogram of magnetic field angles observed using the SMA \citep{sabin2014} and CARMA \citep{sabin2015} are shown in cyan and yellow, respectively.  The mean \textit{Planck} magnetic field angle, {$19.0^{\circ}\pm 3.9^{\circ}$ E of N}, is shown as a red line, with its standard deviation shown as a shaded red sector.  The angle of the outflow from OH231.8, $21^{\circ}$ E of N \citep{alcolea2001}, is shown as a solid black line, and the normal to the outflow is shown as a dashed black line.  The left-hand side of the plot shows the POL-2 magnetic field histogram only.  The right-hand side of the plot overplots the POL-2, SMA and CARMA data.  Plotting and normalization are as described in Figure~\ref{fig:histogram_crl618}. }
    \label{fig:histogram_oh231}
\end{figure}

The average RMS noise in the central 3 arcmin of the map on default 4-arcsec pixels is 6.0 mJy\,beam$^{-1}$ (Stokes $I$), 6.1 mJy\,beam$^{-1}$ (Stokes $Q$), and 6.3 mJy\,beam$^{-1}$ (Stokes $U$) for CRL 618, and 7.5 mJy\,beam$^{-1}$ (Stokes $I$), 7.2 mJy\,beam$^{-1}$ (Stokes $Q$), and 7.2 mJy\,beam$^{-1}$ (Stokes $U$) for OH231.8+4.2.

We binned our output vector catalogues to 8-arcsec (approximately Nyquist-sampled) pixels.  The average RMS noise in the central 3 arcmin of the map on 8-arcsec pixels is 2.0 mJy\,beam$^{-1}$ (Stokes I), 2.0 mJy/beam (Stokes $Q$) and 2.1 mJy/beam (Stokes $U$) for CRL 618, and 2.5 mJy\,beam$^{-1}$ (Stokes I), 2.5 mJy/beam (Stokes $Q$) and 2.5 mJy/beam (Stokes $U$) for OH231.8.  All of the following analysis is performed on these 8-arcsec pixel images.

The observed polarized intensity is given by
\begin{equation}
    P^{\prime} = \sqrt{Q^{2} + U^{2}}.
    \label{eq:poli}
\end{equation}
We debiased this quantity using the modified asymptotic estimator \citep{plaszczynski2014,montier2015}:
\begin{equation}
    P = P^{\prime} - \frac{1}{2}\frac{\sigma^{2}}{P^{\prime}}\left(1-e^{-\left(\frac{P^{\prime}}{\sigma}\right)^{2}}\right),
    \label{eq:poli_debias}
\end{equation}
where $\sigma^{2}$ is the weighted mean of the variances {$\sigma_{Q}^{2}$ and $\sigma_{U}^{2}$},
\begin{equation}
    \sigma^{2} = \frac{Q^{2}\sigma_{Q}^{2} + U^{2}\sigma_{U}^{2}}{Q^{2} + U^{2}},
\end{equation}
calculated on a pixel-by-pixel basis.
Debiased polarization fraction is given by 
\begin{equation}
    p = \frac{P}{I},
\end{equation}
and polarization angle is given by
\begin{equation}
    \theta_{p} = 0.5\arctan\left(\frac{U}{Q}\right).
\end{equation}
We note that the polarisation directions which we detect are not true vectors, as they occupy a range in angle $0-180^{\circ}$ only.  We nonetheless refer to our measurements as vectors for convenience, in keeping with the general convention in the field.  Throughout this work, we assume that magnetic field direction can be inferred by rotating polarization angles by 90$^{\circ}$ \citep{andersson2015}.

The resolution of these observations is 14.1$^{\prime\prime}$ \citep{dempsey2013}.  This is equivalent to 12\,700 AU at the distance of CRL 618, and 21\,700 AU at the distance of OH231.8.  Neither source is well-resolved in our observations.

\section{Results}
\label{sec:results}

Figures~\ref{fig:crl618} and \ref{fig:oh231} show the polarization and magnetic field vector maps for CRL 618 and OH231.8, respectively.  In these figures, and throughout the following analysis, we use the vector selection criteria $I/\delta I > 5$ and $\delta\theta < 9.55^{\circ}$, where the latter criterion is equivalent to a SNR in polarization fraction of $p/\delta p > 3$ \citep{serkowski1962}.

\subsection{Magnetic field morphology}
\label{sec:mag_field_morph}

\begin{figure}
    \centering
    \includegraphics[width=0.47\textwidth]{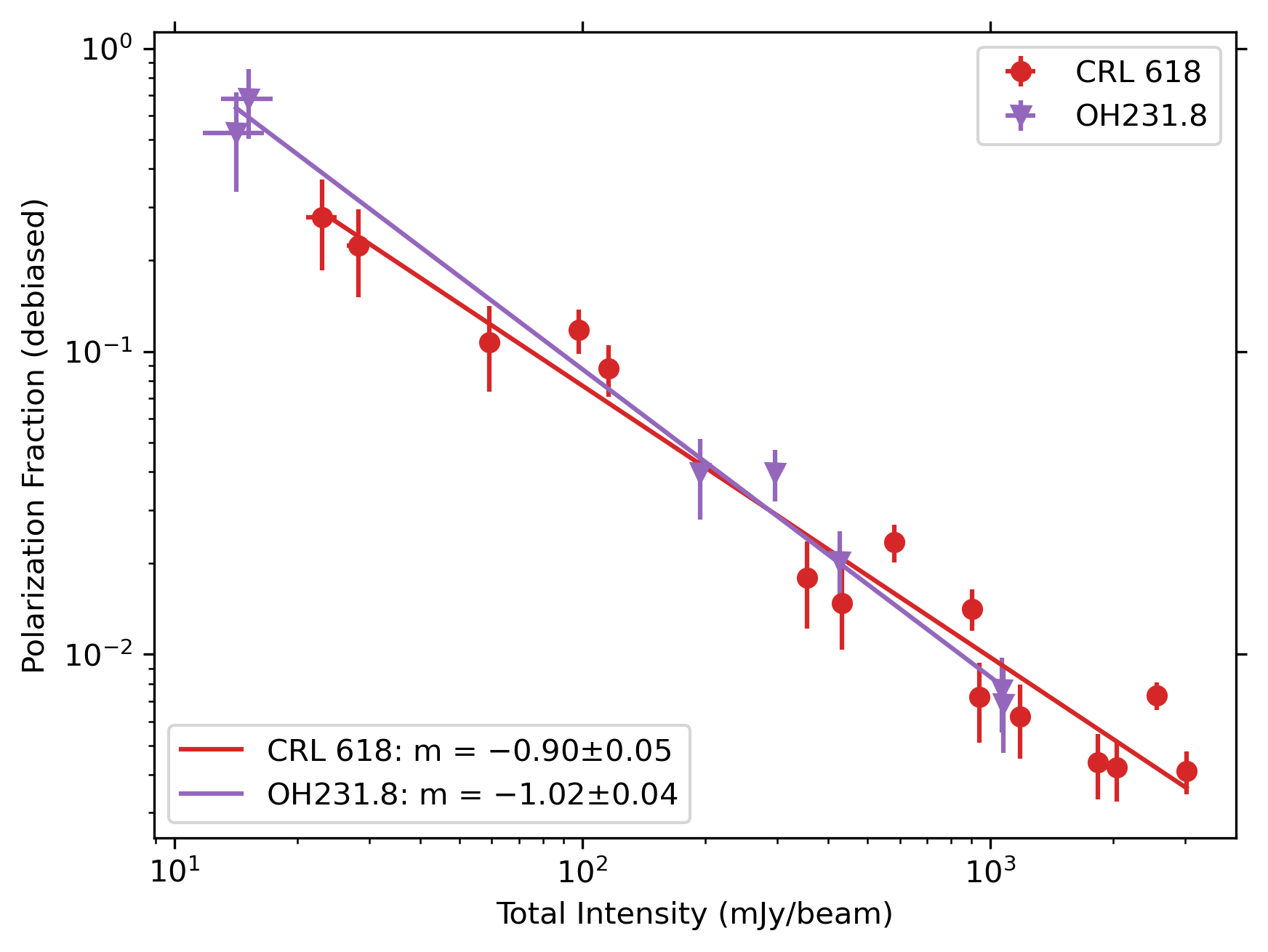}
    \caption{Debiased polarization fraction as a function of total intensity (Stokes $I$) for CRL 618 (red circles) and OH231.8 (purple triangles).  Only data points with $I/\delta I > 5$ and $\delta\theta < 9.55^{\circ}$ (equivalent to $p/\delta p > 3$) are shown.}
    \label{fig:polfrac}
\end{figure}

While neither source is well-resolved in our observations, both show ordered magnetic field geometries.

Figure~\ref{fig:crl618} shows that CRL 618 has a broadly linear field across its centre, oriented $66^{\circ}\pm12^{\circ}$ east of north (mean angle measured over the ten central vectors in the source).  The three vectors to the north and east of the source have a mean magnetic field angle of $161^{\circ}\pm 5^{\circ}$, while the two vectors in the south of the source have a mean angle of $33^{\circ}$.  A histogram of magnetic field angles for CRL 618 is shown in Figure~\ref{fig:histogram_crl618}.

Figure~\ref{fig:oh231} shows that we have a better detection of the magnetic field in the south of OH231.8 than in the north.  The five vectors detected in the centre of the source have an average magnetic field angle of $108^{\circ}\pm 18^{\circ}$ east of north, while the two vectors detected in the western periphery of the source have an average magnetic field angle of $15^{\circ}$.  A histogram of magnetic field angles for OH231.8 is shown in Figure~\ref{fig:histogram_oh231}.

\subsection{Polarization fraction}
\label{sec:polfrac}

\begin{figure}
    \centering
    \includegraphics[width=0.47\textwidth]{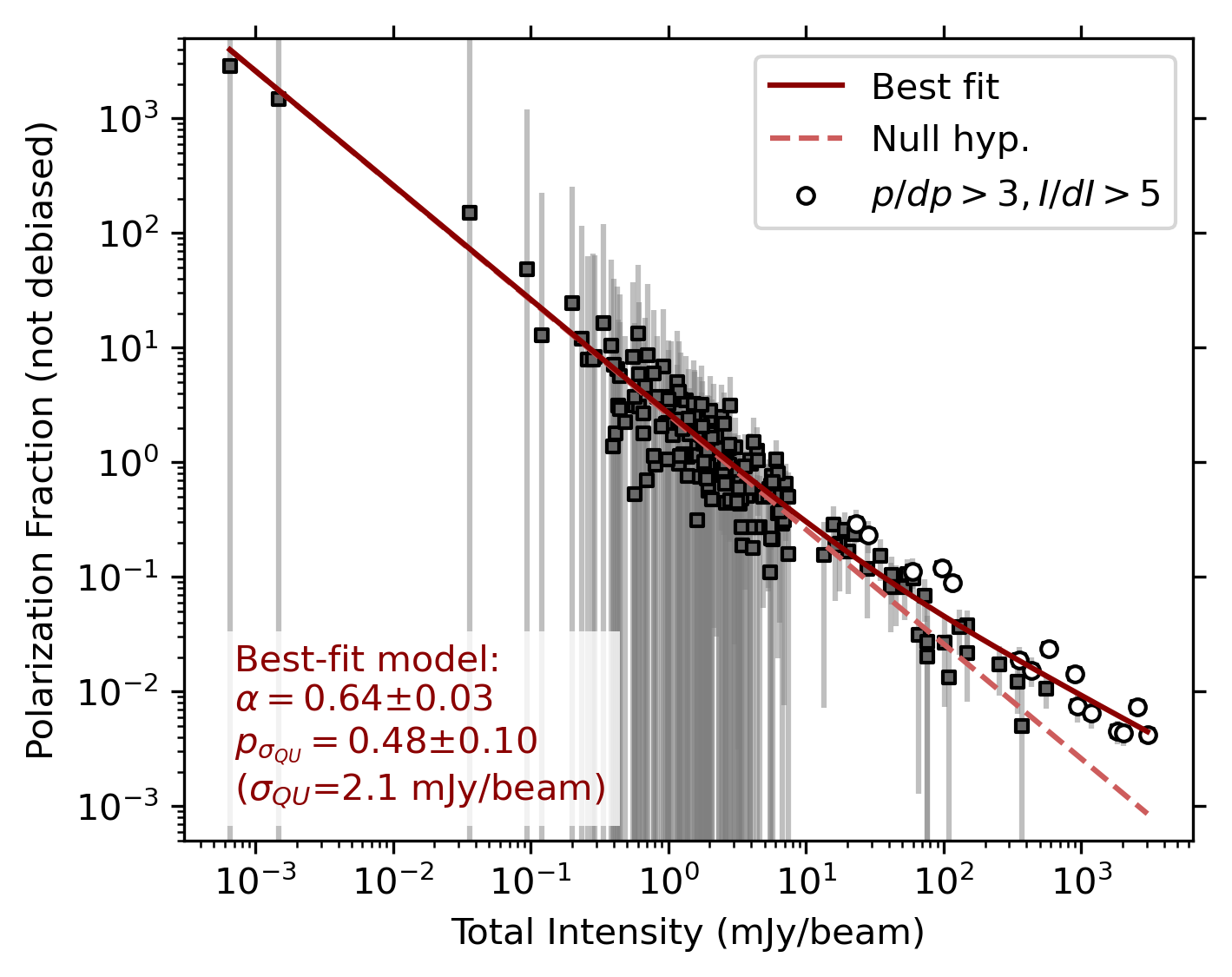}
    \caption{Non-debiased polarization fraction as a function of total intensity for CRL 618. The data are fitted with a single-power-law distribution and a Ricean noise model, as described in the text.  All data points in the central 3-arcmin-diameter region of the image are shown and fitted; those with {$p/\delta p > 3$ and $I/\delta I >5$ are shown as large white circles, while the remainder of the data points are shown as small grey squares}.  The best-fit model is shown as a solid maroon line.  The behaviour in the absence of true polarised signal, $p^{\prime} = \sqrt{\pi/2}(I/\sigma_{QU})^{-1}$, is shown as a dashed {light red} line.}
    \label{fig:pI_crl618}
    \includegraphics[width=0.47\textwidth]{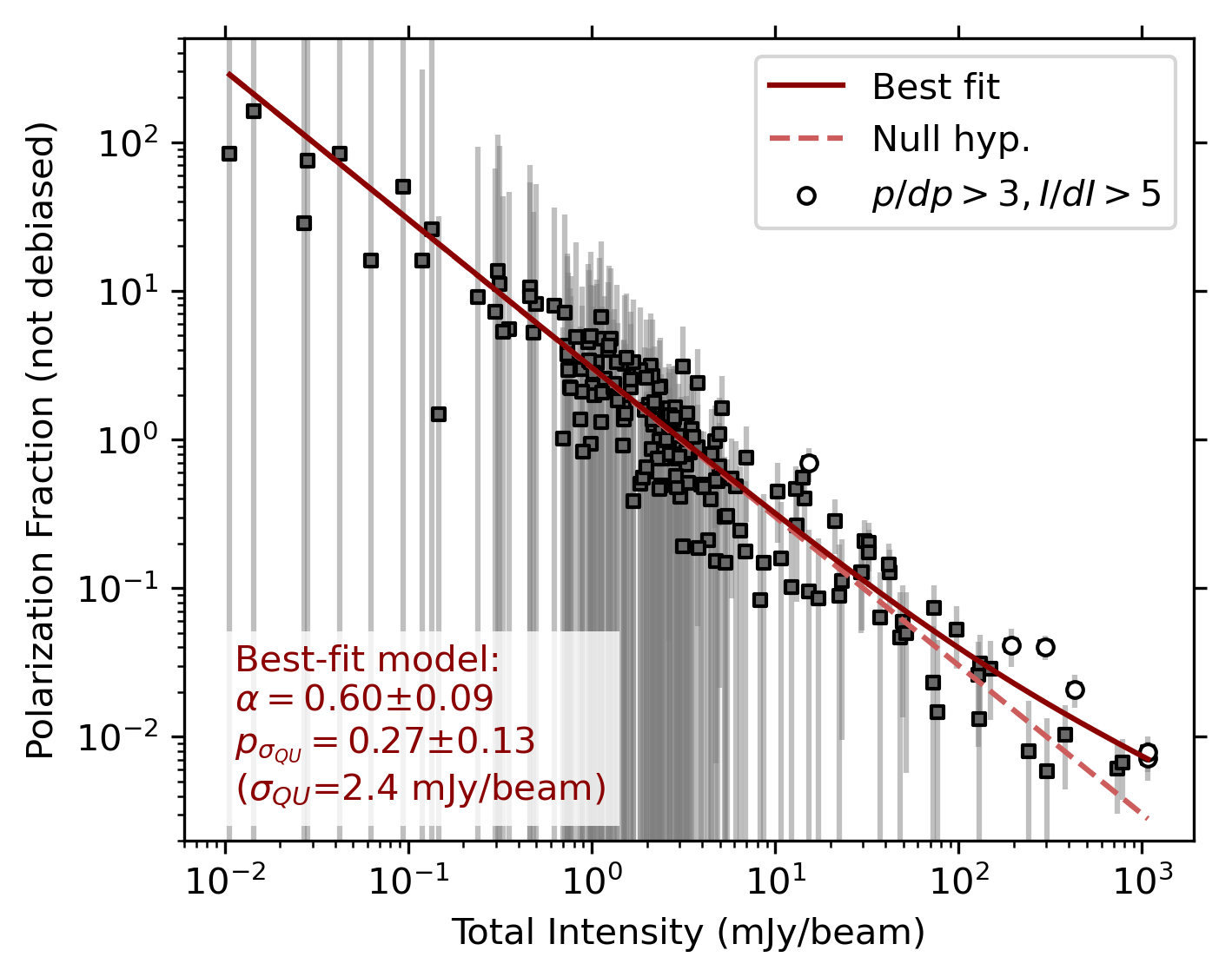}
    \caption{Non-debiased polarization fraction as a function of total intensity for OH231.8.  The features of the figure are as described in Figure~\ref{fig:pI_crl618}.}
    \label{fig:pI_oh231}
\end{figure}

{Dust polarization observations made with single-dish instruments very commonly show a decrease in polarization fraction with increasing source brightness \citep[e.g.][]{goodman1992,wardthompson2000,whittet2008,alves2014,jones2015,pattle2019a}.  This behaviour, commonly referred to as the `polarization hole' effect, is seen both in extinction and emission polarization observations \citep[e.g.][]{alves2014}, and is generally thought to arise from some combination of loss of grain alignment at high extinction and integration of complex small-scale magnetic fields within the telescope beam \citep[e.g.][]{hoang2021}.}  

The polarization fraction of dust emission typically shows a power-law dependence on total intensity, such that 
\begin{equation}
p\propto I^{-\alpha},
\label{eq:p_vs_I}
\end{equation}
where $0\leq \alpha \leq 1$ \citep{whittet2008,jones2015}; {note that total intensity is a proxy for $A_{V}$ in optically thin, isothermal material \citep{pattle2019a}}.  In the Radiative Torques (RAT) paradigm of grain alignment \citep{lazarian2007}, a steeper index indicates either poorer grain alignment with respect to the magnetic field or more variation of the magnetic field direction along the line of sight (LOS).  Thus, in equation~\ref{eq:p_vs_I}, $\alpha = 0$ indicates that grains are consistently aligned throughout the LOS, while $\alpha = 1$ implies complete randomization of either grain alignment or magnetic field direction along the LOS \citep{jones2015,pattle2019a}.

{Attempts to measure $\alpha$ are complicated by the defined-positive nature of polarized intensity (equation~\ref{eq:poli}), which results in a strong positive bias in the measured polarization fraction at low total intensities \citep{serkowski1958}.  Correction of measured polarized intensity values for this effect are known as (statistical) debiasing \citep{wardle1974}.  As discussed above, we use the modified asymptotic estimator method \citep{plaszczynski2014,montier2015} in this work.  However, even after debiasing some positive bias typically remains in low-to-intermediate SNR data \citep[e.g.][]{montier2015}, and so the measured value of $\alpha$ may be artificially high, particularly in data sets such as those presented in this work, which have relatively few high-SNR detections.  Because of this, we use two methods to measure $\alpha$: the standard method of fitting a power law to debiased data \citep[e.g.][]{alves2014,alves2015}, and an alternative method which fits the non-debiased data directly \citep{pattle2019a}.}

\subsubsection{Power-law fit to debiased data}

Figure~\ref{fig:polfrac} shows the relationship between debiased polarization fraction $p$ and total intensity $I$ for CRL 618 and OH231.8, for vectors meeting our selection criteria of $I/\delta I > 5$ and $\delta\theta > 9.55^{\circ}$.  There is no significant difference in the magnitude or distribution of the polarization fractions of the two sources when they are observed with POL-2.  A two-sided KS test gives a 62\% probability that the two samples are drawn from the same underlying distribution.

We performed linear regressions on $\log_{10}p$ as a function of $\log_{10}I$ for each of our sources.  The lines of best fit are shown on figure~\ref{fig:polfrac}.  For both sources, we see a steep negative power-law relationship, with $\alpha = 0.90\pm0.05$ in CRl 618 and $\alpha = 1.02\pm 0.04$ in OH231.8.  This na\:ively implies that there is either little or no grain alignment, or no order to the magnetic field, in either source.  However, the consistent polarization angles that we see in both regions are indicative of an ordered magnetic field being traced by aligned grains {\citep{naghizadehkhouei1993}}.

\subsubsection{Ricean-mean fitting to non-debiased data}

{The} simple linear regression method {presented above} does not take into account the non-Gaussian noise properties of polarized intensity, $P$: assuming that Stokes $Q$ and $U$ have independent Gaussian measurement uncertainties, $P$ will be Ricean-distributed \citep{wardle1974,simmons1985}.  To {attempt to} account for this effect, we {also used an alternative method to measure $\alpha$ \citep{pattle2019a}.  In this approach, we} assume that the underlying relationship between $p$ and $I$ can be parametrized as
\begin{equation}
    p = p_{\sigma_{QU}}\left(\frac{I}{\sigma_{QU}}\right)^{-\alpha}
    \label{eq:polfrac}
\end{equation}
where $p_{\sigma_{QU}}$ is the polarization fraction at the RMS noise level of the data $\sigma_{QU}$, and $\alpha$ is a power-law index in the range $0 \leq \alpha \leq 1$.  {(Note that this is a restatement of equation~\ref{eq:p_vs_I}, parametrized in terms of $\sigma_{QU}$ and $p_{\sigma_{QU}}$.)}

We fitted the relationship between $I$ and observed non-debiased polarisation fraction $p^{\prime}$ with the mean of the Ricean distribution of observed values of $p$ which would arise from equation~\ref{eq:polfrac} in the presence of Gaussian RMS noise $\sigma_{QU}$ in Stokes $Q$ and $U$:
\begin{equation}
    p^{\prime}(I) = \sqrt{\frac{\pi}{2}}\left(\frac{I}{\sigma_{QU}}\right)^{-1}\mathcal{L}_{\frac{1}{2}}\left(-\frac{p_{\sigma_{QU}}^{2}}{2}\left(\frac{I}{\sigma_{QU}}\right)^{2(1-\alpha)}\right).
    \label{eq:rmfit}
\end{equation}
where $\mathcal{L}_{\frac{1}{2}}$ is a Laguerre polynomial of order $\frac{1}{2}$.  See \citet{pattle2019a} for a derivation of this result.  {This model has two regimes; a low-SNR regime in which the data points are characterised by the Ricean noise relation,}
\begin{equation}
    p^{\prime}(I) = \sqrt{\frac{\pi}{2}}\left(\frac{I}{\sigma_{QU}}\right)^{-1},
    \label{eq:null_hyp}
\end{equation}
{and a high-SNR regime in which $p^{\prime}\approx p$, and the data points are characterised by equation~\ref{eq:polfrac}.}

We restricted our data set to the central 3-arcminute diameter region over which exposure time, and so RMS noise, is approximately constant \citep{friberg2016}.  {All data points within this central region are included in the fit, although the data are weighted by their uncertainties, so low-SNR data points have very little effect on the best-fit values.}

The relationship between $p^{\prime}$ and $I$ in CRL 618 is shown in Figure~\ref{fig:pI_crl618}, and for OH231.8 in Figure~\ref{fig:pI_oh231}.  {We note that since the data have intentionally not been debiased, many points with unphysically large polarization fractions appear in the low-SNR regions of these plots.  Since we are modelling both the well-characterised data points and the statistical noise in the data, all of the data points are shown.  Both plots show that equation~\ref{eq:null_hyp} is a good model for the low-SNR data, but not for the high-SNR data points.}

By fitting Equation~\ref{eq:rmfit} to the data, we measure a best-fit index of $0.64\pm 0.03$ in CRL 618 and $0.60\pm 0.09$ in OH231.8, significantly shallower than that determined by performing a {naive} linear regression on the debiased data.

\subsubsection{Interpretation}

{The power-law fit and Ricean-mean methods thus produce quite different values of $\alpha$.  \citet{pattle2019a} found that for true values of $\alpha$ in the range $0.6\lesssim \alpha \lesssim 1$, the power-law fitting method is likely to overestimate $\alpha$, while the Ricean-mean method is likely to underestimate $\alpha$.  The discrepant values that we measure using the two methods thus suggest that the true values of $\alpha$ are likely in the range $0.65-0.9$ for CRL618 and $0.6-1.0$ for OH231.8.}

{These relatively high values of $\alpha$ suggest that there is either loss of grain alignment or field tangling in both of these sources, which is creating significant polarization holes in both sources.  The fact that there is a strong detection of complex polarized emission on smaller scales in interferometric observations (see Sections~\ref{sec:crl618_sma} and \ref{sec:oh231_sma}, below) supports the latter of these hypotheses.  However, both the values of $\alpha < 1$ and the well-ordered polarization angles in both sources indicate that out observations are providing information on the magnetic field direction on $\sim 10^{4}$-au size scales in both sources.}

{We discuss the statistical similarity of the polarization fractions in the two sources in Section~\ref{sec:discuss_polfrac}, below.}

\section{Discussion}
\label{sec:discussion}

\begin{figure}
    \centering
    \includegraphics[width=0.47\textwidth]{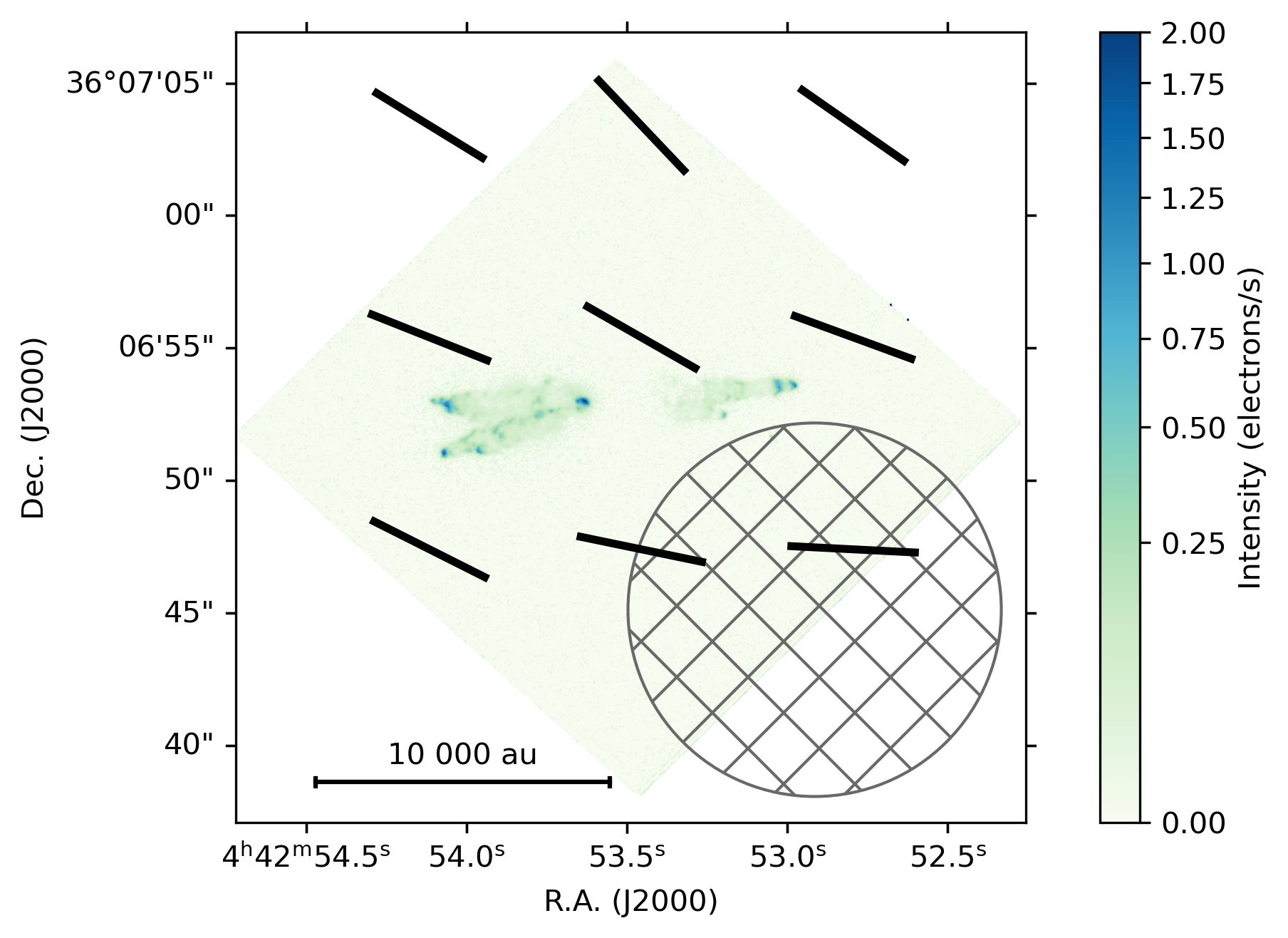}
    \caption{POL-2 magnetic field vectors overlaid on an HST 658nm image of CRL 618.  The JCMT beam size is shown as a hatched circle in the lower right-hand corner of the plot.}
    \label{fig:crl618_hst}
\end{figure}

\subsection{CRL 618}

\begin{figure}
    \centering
    \includegraphics[width=0.47\textwidth]{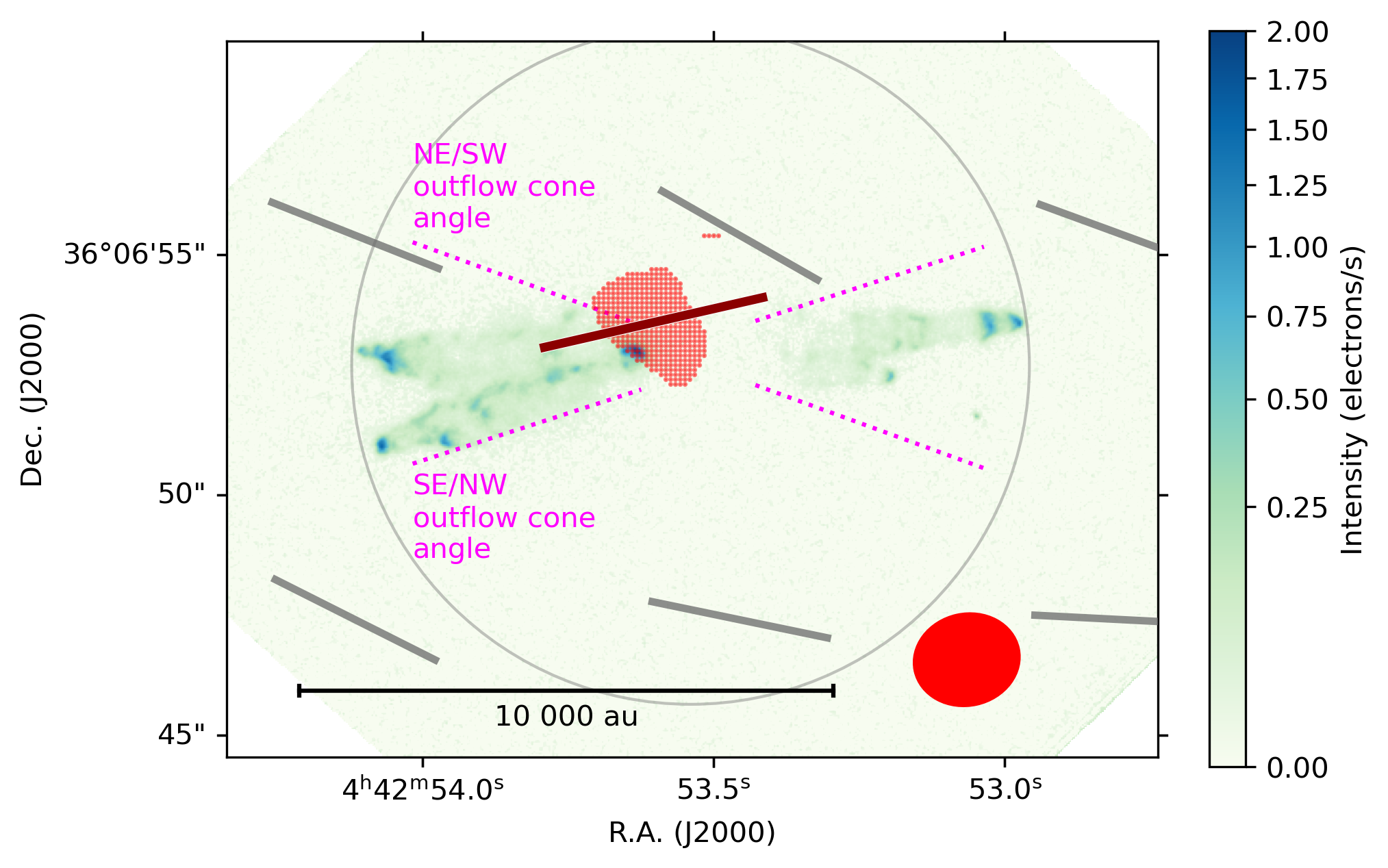}
    \caption{POL-2 magnetic field vectors (grey) and SMA magnetic field vectors (red), overlaid on an HST 658nm image of OH231.8.  For the SMA data, the positions of each vector are shown as dots, and the average magnetic field direction is shown as a vector.  The most extreme position angles of the bullets in the explosive outflow (the `outflow cone angles') are marked with dotted magenta lines.  The JCMT and SMA beams are shown as an open grey circle and a filled red ellipse, respectively.}
    \label{fig:crl618_sma}
\end{figure}

In CRL 618, the magnetic field that we observe in the centre of the source has a position angle {of $66^{\circ}\pm12^{\circ}$,} very similar to the {one of the extreme position angles of bullet ejections from the central source, $70^{\circ}$ E of N.  This angle is hereafter referred to as} the NE/SW opening angle of the dynamical outflow from {CRL 618}.  {The magnetic field in the centre of CRL 618 is significantly ($65^{\circ}$) offset from the $131^{\circ}$ E of N large-scale magnetic field direction observed by \textit{Planck}, as shown in Figures~\ref{fig:crl618} and \ref{fig:histogram_crl618}.} 

The magnetic field in the periphery of CRL 618 differs from both the field in the centre of the source and from the field observed by \textit{Planck}.  The field in the northern and eastern periphery of {CRL 618} is oriented $162^{\circ}\pm5^{\circ}$ E of N, approximately perpendicular to the outflow direction, and similar to (offset $\simeq 30^{\circ}$ from) the large-scale magnetic field direction in the region, {$131^{\circ}$ E of N}, as measured from \textit{Planck} Observatory dust polarization data\footnote{\url{https://www.cosmos.esa.int/web/planck/pla}} (see Figure~\ref{fig:crl618}) with a resolution of $5^{\prime}$ \citep{planck2014xi}.  The magnetic field in the southern periphery of the source is oriented $33^{\circ}$ E of N, more similar to (offset $\simeq 30^{\circ}$ from) the magnetic field direction observed in the centre of the source than to the field observed by \textit{Planck}, to which it is almost perpendicular (offset by $82^{\circ}$).

These deviations of the magnetic field in CRL 618 from that of the large-scale field traced by \textit{Planck} indicate either that the magnetic field in CRL 618 is unrelated to that in the dust traced by \textit{Planck} along the same line of sight, or that the magnetic field within CRL 618 has been significantly reordered, potentially by the outflow from the central source.

\subsubsection{Comparison with HST imaging}

Figure~\ref{fig:crl618_hst} shows our POL-2 magnetic field vectors overplotted on Hubble Space Telescope (HST) imaging of CRL 618.  This data was taken on 2009 Aug 08 with the Wide Field Camera 3 (WFC3) UVIS1 detector, using the F658N narrow-band filter.  The data, first published by \citet{balick2013}, were retrieved from the Hubble Legacy Archive\footnote{\url{https://hla.stsci.edu}; Project ID 11580}.  This filter covers the [N \textsc{ii}] forbidden line at 658.3\,nm, a tracer of high-velocity material in jets \citep{nishikawa2008}.  Although CRL 618 has been observed using a wide range of HST filters, we chose F658N both as a high-velocity emission tracer, and for consistency with OH231.8, which has only been observed using the F658N and F606W wide V-band filters.

The HST 658\,nm imaging traces the explosive bullets emanating from the post-AGB star at the centre of CRL 618, as shown in Figure~\ref{fig:crl618_hst}.  This outflow is not resolved in our POL-2 observations.  The POL-2 vectors do not align with the average direction of the bullets, instead having an average position angle similar to the NE/SW opening angle of the outflow.

\subsubsection{Comparison with SMA magnetic field}
\label{sec:crl618_sma}

Figure~\ref{fig:crl618_sma} overplots both POL-2 and Submillimeter Array (SMA) 850\,$\mu$m magnetic field vectors on the HST 658\,nm image of CRL 618.  The SMA observations \citep{sabin2014} have a beam size of $2\farcs2\times 1\farcs9$ and a beam position angle of $-77.6^{\circ}$, and cover an area of $\sim 5\farcs4 \times 4\farcs6$ \citep{sabin2014} in the centre of CRL 618, approximately 9\% of the area of the JCMT beam (cf. \citealt{dempsey2013}), with polarized emission detected over the central few square arcseconds of the region over which total emission is detected.  There is a clear disagreement between the polarization angles measured by the SMA and the JCMT.  The SMA vectors have a mean magnetic field angle of $97^{\circ}\pm 17^{\circ}$, compared to a mean magnetic field angle of $66^{\circ}\pm 12^{\circ}$ over the central POL-2 vectors in the source (see Section~\ref{sec:mag_field_morph}).  The discrepancy between the JCMT and SMA magnetic field distributions is shown in Figure~\ref{fig:histogram_crl618}.

\subsubsection{Interpretation}

Neither our data nor the SMA observations have mean position angles that match the average position angle of the outflowing bullets ($89^{\circ}$ E of N; \citealt{huang2016}).  Instead, both appear to coincide with one of the two extreme position angles of the collection of fast molecular outflows (``bullets'') that are emanating from the centre of CRL 618 \citep{huang2016}.  These two extreme angles, which we take to give an indication of the likely angular extent of a cavity opened in the dusty CSE by the cone of outflowing bullets, are referred to as `outflow cone angles' in the following discussion, and are marked on Figure~\ref{fig:crl618_sma}.

While our JCMT POL-2 data coincide with the NE/SW outflow cone angle ($70^{\circ}$ E of N; a sector angle of 18$^{\circ}$ about the outflow cone mean position angle of 88$^{\circ}$ E of N), the SMA data, which trace much smaller scales, appear to coincide with the SW/NE opening angle ($106^{\circ}$ E of N; \citealt{huang2016}).  This is consistent with our finding that polarization fraction decreases with increasing polarized intensity in CRL 618 (cf. Section~\ref{sec:polfrac}), as integration over a more complex small-scale magnetic field in the vicinity of the central source will result in a decrease in polarization fraction.

It is not possible to tell from our relatively low-resolution observations where within CRL 618 the polarized emission that we observe preferentially arises from.  One possibility is that the magnetic field direction that we infer accurately represents the typical magnetic field direction across the dusty CSE surrounding V353 Aur generated in the earlier AGB phase of the star's lifetime \citep[e.g.][]{habing2004}.  Alternatively, we can speculate that we are preferentially observing emission arising from heated and compressed CSE material in the northeastern wall of a cavity through the CSE opened by the outflowing cone of bullets.  As the eastern side of the CRL 618 outflow is moving towards us, with the northernmost bullets being located closest to the plane of the sky \citep{huang2016}, the north-eastern cavity wall may be that which is most directly visible to us. It is probable in this case that the magnetic fields have been reordered from their original orientation (possibly indicated by the directions of the vectors in the periphery of the core) to run along the outflow cavity walls.

\begin{figure*}
    \includegraphics[width=0.49\textwidth]{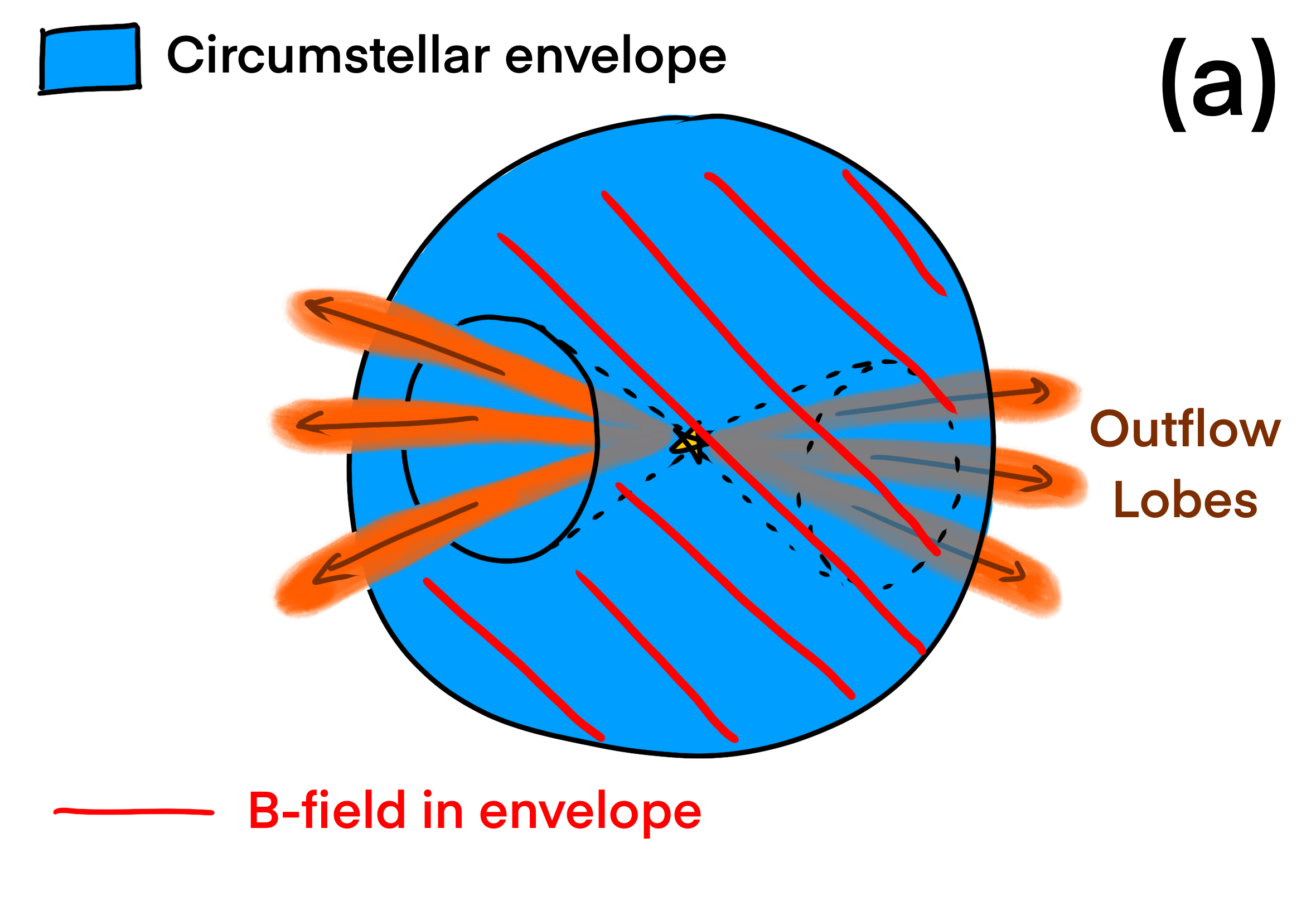}\hfill
    \includegraphics[width=0.49\textwidth]{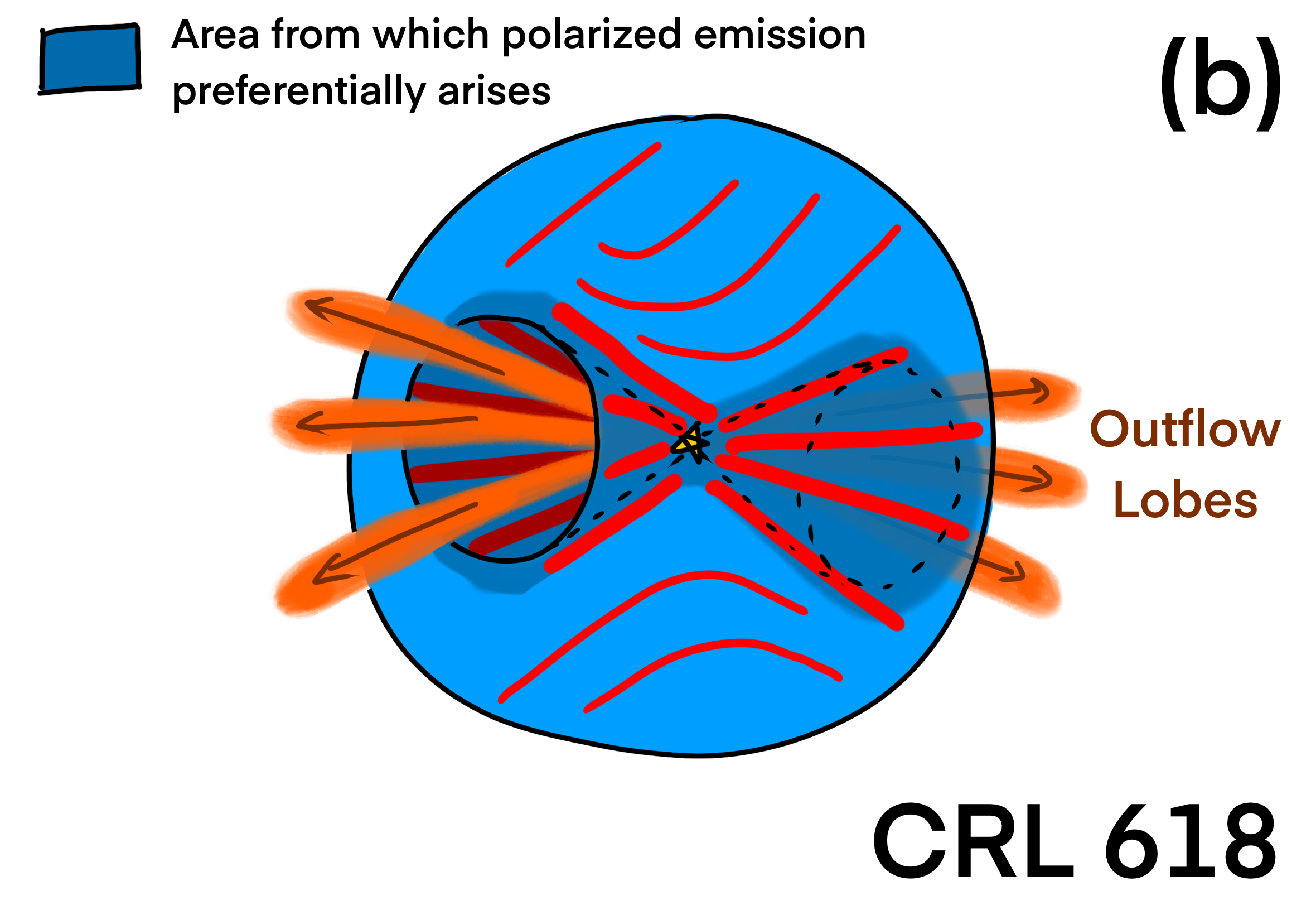}
    \caption{Cartoons illustrating our two hypotheses for the origin of the polarized emission that we observe in CRL 618. 
    \textit{Left:} Our observations trace the magnetic field direction in the bulk of the CSE.  \textit{Right:} Our observations preferentially trace the magnetic field in heated and compressed CSE material in the outflow cavity walls.  This field results from a pre-existing field that has been reorganised by the outflow-CSE interaction to run parallel to the cavity wall, as is commonly seen in outflow cavity walls in protostellar envelopes.}
    \label{fig:crl618_cartoon}
\end{figure*}

Our two hypotheses for the origin of the polarized emission that we observe from CRL 618 are illustrated in Figure~\ref{fig:crl618_cartoon}.  While we cannot distinguish between these two hypotheses with our current low-resolution observations, we note that the latter is in keeping with the observed behaviour of magnetic fields in protostellar envelopes under the influence of outflows, as we discuss further in Section~\ref{sec:discuss_outflows}, below.

\subsection{OH231.8+4.2}

\begin{figure}
    \centering
    \includegraphics[width=0.47\textwidth]{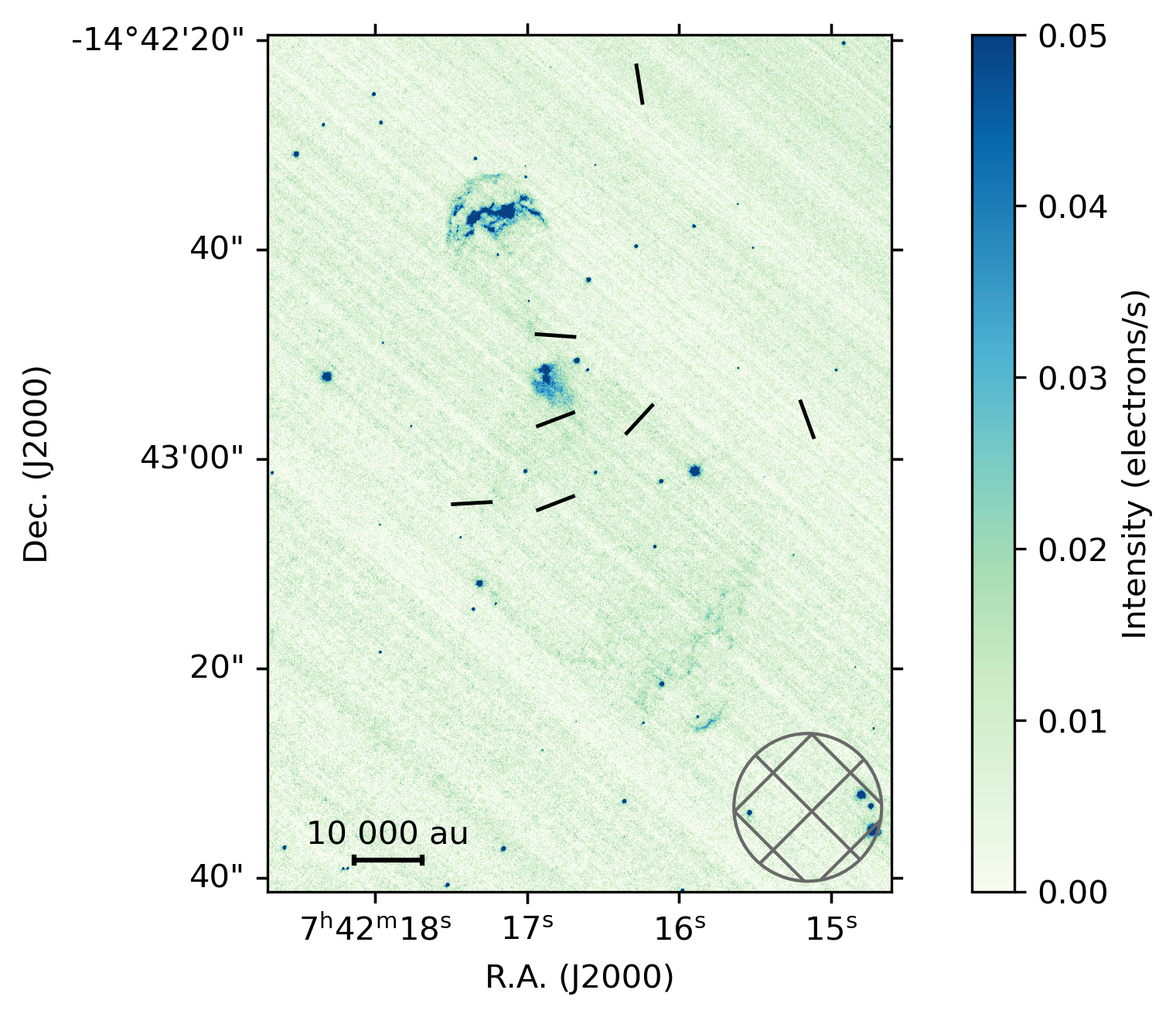}
    \caption{POL-2 magnetic field vectors overlaid on an HST 658nm image of OH231.8.  The JCMT beam size is shown as a hatched circle in the lower right-hand corner of the plot.}
    \label{fig:oh231_hst}
\end{figure}

\begin{figure*}
    \centering
    h\includegraphics[width=\textwidth]{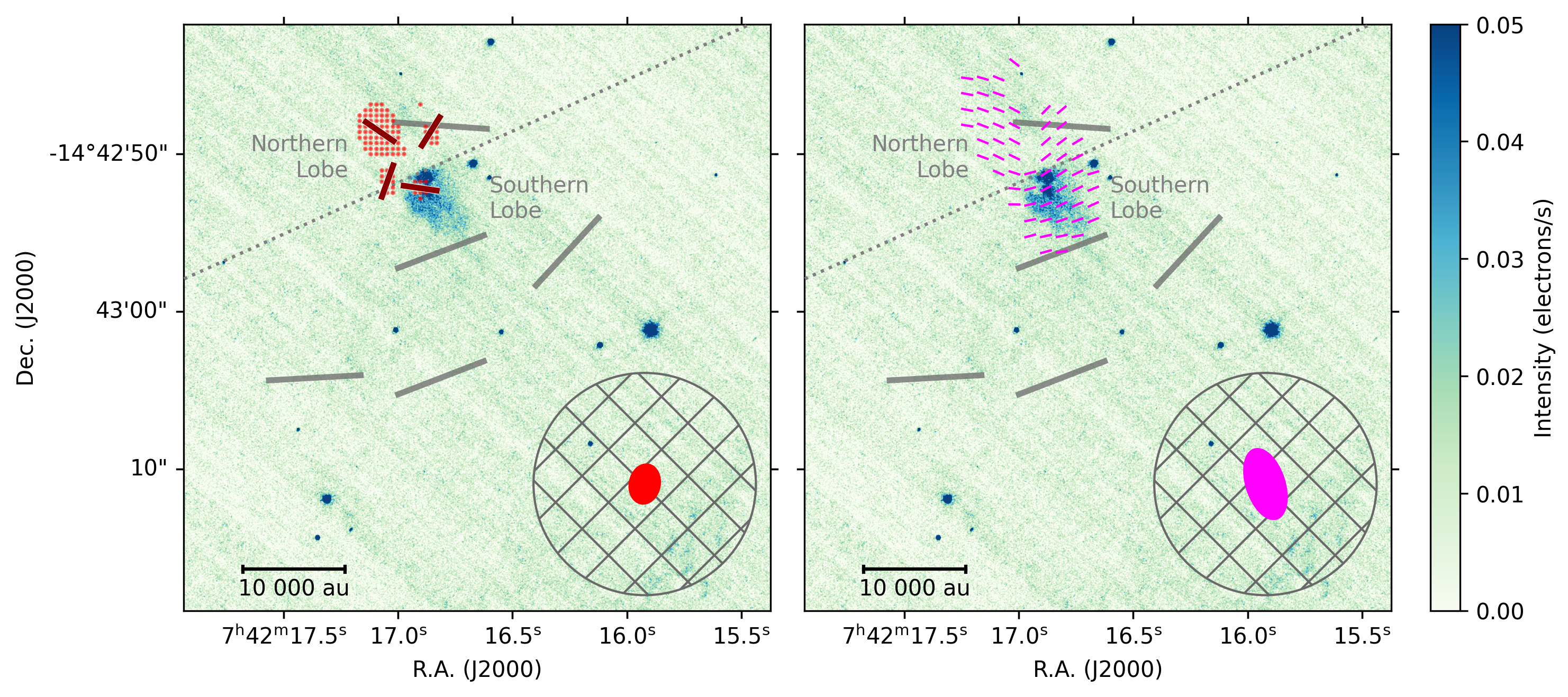}
    \caption{POL-2 magnetic field vectors (grey) overlaid on an HST 658nm image of OH231.8.  \textit{Left:} with SMA magnetic field vectors \citep{sabin2014}.  For the SMA data, the positions of each vector are shown as dots, and the average magnetic field directions of each of the four contiguous regions where polarization is detected are shown as vectors.  \textit{Right:} with CARMA magnetic field vectors overlaid \citep[magenta, 1 in 16 vectors shown, for clarity]{sabin2015}.  The division between the northern and southern lobes is shown as a grey dotted line.  In both panels the JCMT beam size is shown as a hatched area in the right-hand corner.  The SMA and CARMA beams are shown as red and magenta ellipses in the left- and right-hand panels, respectively.}
    \label{fig:oh231_sma}
\end{figure*}

In OH231.8, we preferentially detect polarization in the south of the source.  The field is perpendicular to the outflow direction (outflow position angle: 21$^{\circ}$; \citealt{alcolea2001}), albeit also showing some (marginally resolved) signs of curvature in vectors located away from the outflow axis.  In the periphery of the source, the magnetic field direction is approximately parallel both to the outflow direction and to the magnetic field direction observed on large scales by \textit{Planck} (average \textit{Planck} magnetic field direction: $19^{\circ}$) with a resolution of $5^{\prime}$ \citep{planck2014xi}.

\subsubsection{Comparison with HST imaging}

Figure~\ref{fig:oh231_hst} shows our POL-2 magnetic field vectors overplotted on Hubble Space Telescope [NII] (658.3\,nm) imaging of OH231.8.  These data were taken on 2009 Dec 13 with WFC3/UVIS1, using the F658N filter, and  were retrieved from the Hubble Legacy Archive\footnote{\url{https://hla.stsci.edu}; Project ID 11580}; data from this project on OH231 were published by \citet{balick2017}, although the F658N map is not shown in that work.  We note that WFPC2 F656N observations of this source were previously published by \citet{bujarrabal2002}. 

The HST 658\,nm emission traces the shocked edges of the bipolar outflow, as well as an area of bright emission close to the centre of the {PPN}.  Our magnetic field vectors are broadly perpendicular to the edges of the outflow, as traced by [NII] emission.

The area of 658\,nm-bright emission near the centre of OH231.8, designated as 2MASS J07421687-1442521, is also bright in visible light, arising from reflection nebulosity, and in the infrared, arising from dust emission {\citep{sanchezcontreras2000,bujarrabal2002}}.  \citet{bujarrabal2002} identify this emission (their Source I$_2$), as dense ($\sim 10^{5}$\,cm$^{-3}$), cool ($\sim 25\,$K) material associated with the collimated outflow from the central source.

\subsubsection{Comparison with SMA and CARMA data}
\label{sec:oh231_sma}

The left-hand panel of Figure~\ref{fig:oh231_sma} overplots POL-2 and SMA 850\,$\mu$m magnetic field vectors on the HST 658\,nm image of OH231.8.  The right-hand panel shows the CARMA 1.3\,mm magnetic field vectors.

As discussed above, while the magnetic field in the southern lobe of OH231.8 is well-detected, in the northern lobe we only detect significant polarization in one $8\times 8$ arcsec$^2$ pixel. (The dividing line between the northern and southern lobes is marked on the right-hand panel of Figure~\ref{fig:oh231_sma}.)  This poorer detection in the northern lobe is consistent with the SMA and CARMA observations, which show a V-shaped magnetic field in the north of the source \citep{sabin2014,sabin2015}, which if integrated over a JCMT beam should produce little or no net polarization.  The magnetic field angle that we measure with POL-2 is consistent with the average magnetic field angle observed with CARMA: our POL-2 vector has a magnetic field angle of 86$^{\circ}$, while the average CARMA magnetic field angle in the northern lobe is $86^{\circ}\pm 28^{\circ}$.  This interpretation is also supported by the low polarization fraction that we measure in our detection in the northern lobe, $0.8\%\pm 0.2\%$.  For comparison, the four pixels in the southern lobe have a mean polarization fraction of 2.7\%$\pm$1.4$\%$, as discussed below.

In the southern lobe of the outflow, where we have a better detection with POL-2 {(with significant polarization detected in four $8\times 8$ arcsec$^2$ pixels)}, the polarization signal is again in good agreement with that seen in the same location by CARMA \citep{sabin2014}.  The mean POL-2 magnetic field angle is $113^{\circ} \pm 16^{\circ}$ in the southern lobe, while the CARMA magnetic field angle is $109^{\circ}\pm 9^{\circ}$.  Both our POL-2 and the CARMA observations show a relatively linear, ordered field, and SMA and ALMA observations also show a coherent magnetic field geometry in the region \citep{sabin2020}.  
The polarized emission seen in both the SMA and CARMA observations of OH231.8's southern lobe appears to preferentially arise from the dusty
clump 2MASS J07421687-1442521. 
This more uniform magnetic field in the southern lobe is reflected both in the better detection and in the higher average polarization fraction in the southern lobe compared to the northern lobe.  The four southern-lobe detections have polarization fractions of $4.0\%\pm1.2\%$, $4.0\%\pm0.8\%$, $2.0\%\pm0.5\%$, and $0.7\%\pm0.2\%$, with the lowest polarization fraction being that of the detection within one beam width of the central star, which also encompasses some of the more complex magnetic field in the northern lobe.  

\subsubsection{Interpretation}

The field in the northern lobe of the outflow appears to be tracing the outflow cavity walls, although we integrate over this signal when observing with POL-2. 

\begin{figure}
    \centering
    \includegraphics[width=0.8\linewidth]{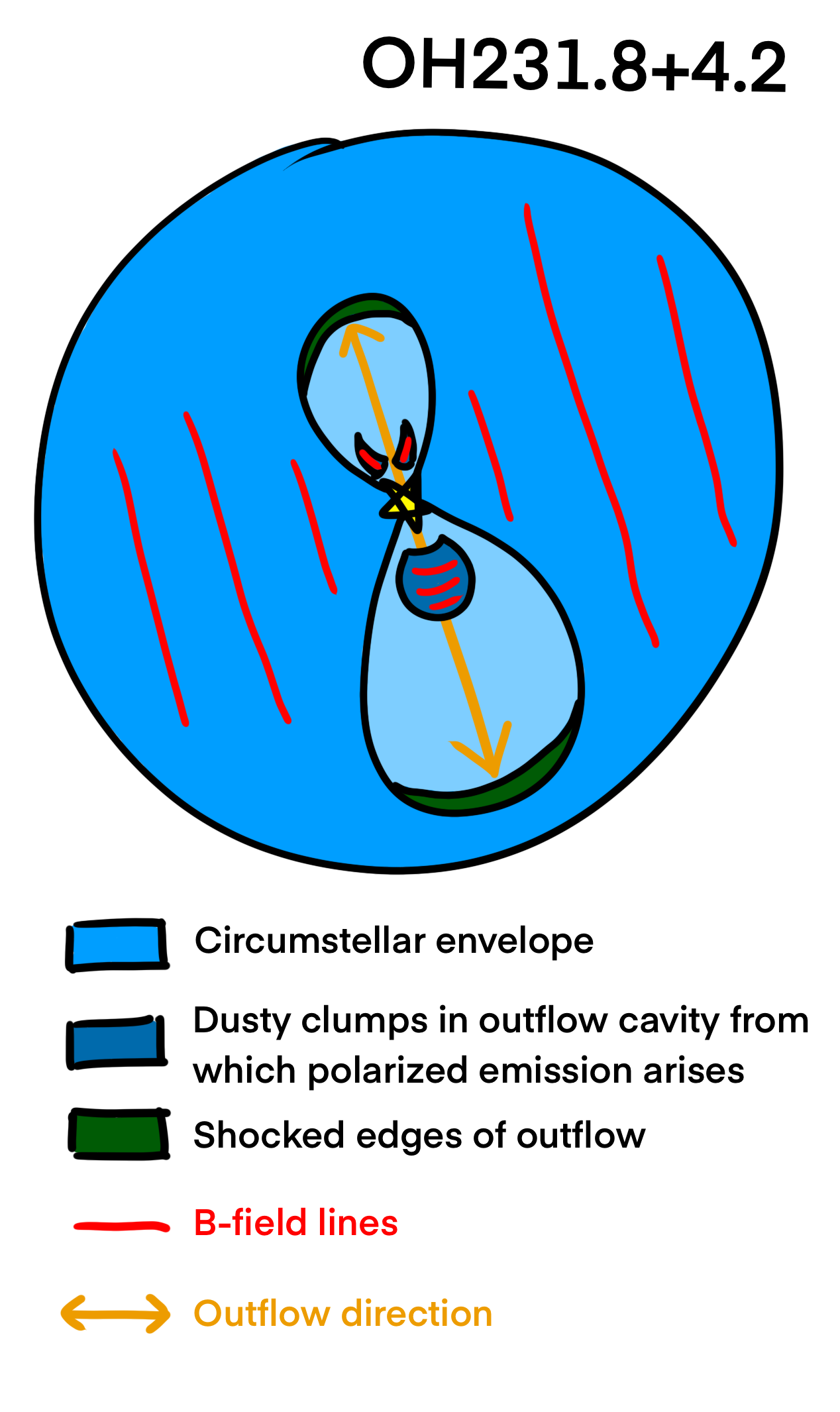}
    \caption{A cartoon illustrating our hypothesis for the origin of the polarized emission that we observe in OH231.8.  The magnetic field in the bulk of the CSE has a direction similar to that of the bipolar outflow.  However, close to the central star, the magnetic field is rearranged by the passage of the outflow.  In the northern outflow lobe, the field traces the outflow cavity walls (this behaviour is seen in the SMA and CARMA data, and integrated over in our POL-2 observations).  In the southern outflow lobe, the magnetic field is perpendicular to the expansion direction of the outflow lobe, and preferentially arises from an IR-bright clump near the base of the outflow.}
    \label{fig:oh231_cartoon}
\end{figure}

The field in the southern lobe is broadly linear, or perhaps perpendicular to the expansion direction of the outflow lobe.  This magnetic field has previously been interpreted to be toroidal around the jet from OH231.8 \citep{sabin2020}.  An interpretation informed by our interpretation of CRL 618, and of the northern lobe of OH231.8, is that the magnetic field that we observe is that which is frozen into circumstellar dust which is being swept up by the {collimated} outflow.  This material is moving away from us along the line of sight \citep[e.g.][]{sabin2014}.  This field appears to be toroidal around the central source, or possibly a field that was initially toroidal that is now being distorted by the passage of the outflow.  The consistency between the CARMA and POL-2 magnetic field angles suggest that both are tracing the same magnetic field, and possibly therefore also that the polarized emission detected by POL-2 preferentially originates from the same material that is detected by CARMA; i.e. from the bright region 2MASS J07421687-1442521.  We illustrate our interpretation of the magnetic field of OH231.8 in Figure~\ref{fig:oh231_cartoon}.

\subsection{The interaction between magnetic fields and outflows}
\label{sec:discuss_outflows}

It appears that the magnetic fields that we observe in dust polarization observations made in the presence of a stellar outflow may preferentially arise from material that has been swept up by the passage of an outflow.  This is consistent with an emerging picture at the other end of the life cycle of stars: of the behaviour of magnetic fields in the vicinity of outflows from protostars.

Examples of magnetic fields in protostellar envelopes which appear to have been reshaped by outflows from their central protostars include Orion B NGC2071IR \citep{lyo2021}, the Bok globule CB54 \citep{pattle2022}, and the low-mass core L43 \citep{karoly2023}.  A similar behaviour has also been seen in the periphery of the wind-blown dust cavity associated with the Herbig Be star HD 200775 \citep{sharma2025}.  In each case, POL-2 dust polarization observations appear to trace magnetic fields in swept-up material in the outflow cavity walls.  \citet{yen2021} performed a statistical analysis of POL-2 observations of protostellar cores, finding that the distribution of angle difference between outflows and local magnetic fields in the plane of the sky peaks at $15^{\circ}-35^{\circ}$, and is consistent with a mean three-dimensional difference in angle of $50^{\circ}\pm15^{\circ}$ in 3D.  This is consistent with the interpretation in which these observation preferentially trace swept-up material in the cavity walls of the outflows, although it should be noted that \citet{yen2021} also could not rule out a random distribution of outflow/magnetic field alignments.

Our results lead us to hypothesize a similar behaviour in CRL 618 and OH231.8, in which we observe magnetic fields in CSE material swept up, compressed, and heated by the passage of outflows.  However, our results show that the observed magnetic field direction depends sensitively on both the geometry of the system and the resolution at which it is observed.  CRL 618 and OH231.8 show effectively opposite relationships between magnetic field and outflow direction when observed with a linear resolution $\sim 10^{4}$\,au, with the magnetic field in CRL 618 being similar to but offset from the outflow direction, while in OH231.8 the magnetic field is broadly perpendicular to the outflow.

In order to accurately probe the magnetic field geometries in the extended dust envelope of AGB stars and PNe, rather than only in the small-scale dust features detected in interferometric imaging, higher resolution single-dish dust polarization observations are required.  Such observations will require the next generation of submillimetre telescopes, such as the Atacama Large-aperture Submillimeter Telescope (AtLAST), which, with an angular resolution of 3.3$^{\prime\prime}$ at 460\,GHz (650$\mu$m), will be able to map PPNe at an angular resolution $\sim 4\times$ better than that of these observations \citep{klaassen2024, mroczkowski2025}.  AtLAST will have an angular resolution at 870\,$\mu$m (345\,GHz) of 4.3$^{\prime\prime}$ \citep{mroczkowski2025}, smaller than the maximum recoverable scales of ALMA in Band 7 (345\,GHz) in its more compact configurations \citep[8.25--4.68$^{\prime\prime}$ in configurations C1--C3,][]{privon2025}. AtLAST total power observations and mosaicked ALMA observations could therefore be feathered \citep{stanimirovic2002,cotton2017} to create high-resolution dust polarization maps of PPNe in which the full range of spatial scales in both the outflows and the circumstellar envelope are recovered.

\subsection{Polarization fraction}
\label{sec:discuss_polfrac}

Interferometric observations of the two sources have suggested that the typical polarization fraction in OH231.8 is higher than that in CRL 618 \citep{sabin2014}, indicative of their differing chemistry and so dust grain composition.  However, as shown in Figure~\ref{fig:polfrac}, there is no significant {statistical} difference in the polarization fractions of the two sources when they are observed with POL-2.  A two-sided KS test gives a 62\% probability that the {debiased} polarization fractions of the high-SNR data points in the two samples {(shown in Figure~\ref{fig:polfrac})} are drawn from the same underlying distribution.  However, as discussed above, the distributions of polarization angles in the two sources are significantly different, and in both cases appear to trace magnetic field geometries associated with outflow structures.

The lack of significant differences in polarization fractions between the two sources, despite their differing chemistry, is likely a result of the differing small-scale magnetic field geometries in the two regions.  In OH231, we are averaging over a complex small-scale magnetic field structure \citep{sabin2014,sabin2020}, and so expect a low net polarization fraction in our single dish observations.  Conversely, the underlying small-scale magnetic field geometry in CRL 618 appears fairly linear, albeit with a notable difference in direction between the scales traced by the SMA and the integrated magnetic field direction that we see with POL-2 \citep{sabin2014}, suggesting that less of the polarization signal is averaged away in our single-dish observations.  It thus appears that at linear resolutions $\sim 10^{4}$\,au, the complexity of the magnetic field geometry on sub-beam scales is the deciding factor in determining the measured polarization fraction of the source, rather than grain composition.

A key tool for investigating the dust polarization properties of PPNe, and AGB star CSEs more generally, will be the Probe far-Infrared Mission for Astrophysics (PRIMA), a proposed polarization-sensitive far-infrared space telescope currently undergoing Phase A study \citep{burgarella2024}.  PRIMA will have polarimetric capability in four broad bands between 91--235\,$\mu$m, with a resolution of $11^{\prime\prime}-27^{\prime\prime}$\footnote{\url{https://prima.ipac.caltech.edu/page/instruments}}, comparable to that of POL-2, but will have a significantly enhanced mapping speed over current ground-based instrumentation. \citet{dowell2024}, modelling the capabilities of PRIMA's polarimetric imager PRIMAger, find that PRIMAger will achieve a sensitivity of 0.0188 MJy/sr over a 10$^{\prime}$-diameter field in 1.5\,h at 235\,$\mu$m.  These numbers are directly comparable to our observations, in which we achieved a sensitivity of 2.5 mJy\,beam$^{-1}$ $= 0.46$\,MJy\,sr$^{-1}$ over the 3$^{\prime}$-diameter central region of our map in 1.6\,h when observing OH231.8, a factor $\sim 270$ decrease in time required to reach the same sensitivity.  However, for dust at $\sim 100$\,K, we might expect a factor $\sim 36$ increase in brightness between 850$\mu$m and 235\,$\mu$m, using the relation $F_{\nu}\propto \kappa_{\nu}B_{\nu}(T)$ \citep{hildebrand1983}, where dust opacity $\kappa_{\nu}\propto\nu^{\beta}$ \citep{beckwith1990}, taking $\beta=0.6$ \citep{murakawa2013}.  This suggests that the SNR achieved in our POL-2 observations could be achieved $\sim 10^{4}$ times faster with PRIMAger.  PRIMA's significantly improved mapping speed over ground-based instrumentation will thus allow both investigation of the peripheries of CSEs, which will be less perturbed by PPNe outflows, and also rapid mapping of a statistically meaningful number of PPNe, thereby determining whether measurable differences in polarization properties arise across the population \citep{sabin2023}, which is not apparent from our observations of two individual sources.

\section{Conclusions}
\label{sec:conclusion}

We have observed the pre-planetary nebulae CRL 618 and OH231.8+4.2 in 850\,$\mu$m polarized light using the POL-2 polarimeter on the SCUBA-2 camera on the James Clerk Maxwell Telescope (JCMT).  In both sources our observations trace magnetic fields which appear to have been shaped by interaction with the outflow from the central stellar system.

CRL 618 shows a magnetic field that is broadly aligned with one of the most extreme position angles of the bullets ejected from the central source, and appears to trace material in the walls of the dust cavity opened by the ejected material.  Conversely, OH231.8 shows a magnetic field that is aligned approximately perpendicular to the outflow direction, and which appears to primarily arise from a hot, IR-bright clump near the base of the outflow.  Our results suggest that in both sources we observe magnetic fields in material from the dusty circumstellar envelope that has been swept up by the passage of outflows from the central star.  However, the observed magnetic field direction depends sensitively on both the geometry of the system and the resolution at which it is observed.

Despite CRL 618 being carbon-rich and OH231.8 being oxygen-rich, there is no significant difference in the polarization fractions of the two sources, likely because in OH231.8 we are integrating over a more complex small-scale magnetic field geometry than is present in CRL 618.  This suggests that at linear resolutions $\sim 10^{4}$\,au, the complexity of the magnetic field geometry on sub-beam scales, rather than grain composition, sets the measured polarization fraction of these sources.  Future polarization-sensitive single-dish instruments will be able to better disentangle the magnetic field and dust polarization properties of PPNe, either through higher-resolution observations with AtLAST, or through rapid mapping of large samples of carbon- and oxygen-rich PPNe with PRIMA.

\section*{Acknowledgements}

The authors thank Laurence Sabin for providing the SMA and CARMA vector catalogues used in this work.

K.P. is a Royal Society University Research Fellow, supported by grant no. URF\textbackslash R1\textbackslash 211322. 

The James Clerk Maxwell Telescope (JCMT) is operated by the East Asian Observatory on behalf of The National Astronomical Observatory of Japan; Academia Sinica Institute of Astronomy and Astrophysics; the Korea Astronomy and Space Science Institute; the National Astronomical Research Institute of Thailand; Center for Astronomical Mega-Science (as well as the National Key R\&D Program of China with No. 2017YFA0402700). Additional funding support is provided by the Science and Technology Facilities Council (STFC) of the United Kingdom and participating universities and organizations in the United Kingdom, Canada and Ireland.  Additional funds for the construction of SCUBA-2 were provided by the Canada Foundation for Innovation.

This research has made use of: Starlink software \citep{currie2014}, currently supported by the East Asian Observatory; Astropy (\url{http://www.astropy.org}), a community-developed core Python package and an ecosystem of tools and resources for astronomy \citep{astropy:2013, astropy:2018, astropy:2022}; the SIMBAD database, operated at CDS, Strasbourg, France; NASA's Astrophysics Data System.

The authors wish to recognize and acknowledge the very significant cultural role and reverence that the summit of Maunakea has always had within the indigenous Hawaiian community.  We are most fortunate to have the opportunity to conduct observations from this mountain.

\section*{Data Availability}

The raw POL-2 data used in this paper are available from the Canadian Astronomy Data Centre (CADC) under project code M23BP031.  The reduced POL-2 data are available at \url{https://dx.doi.org/10.11570/25.0089}.


\bibliographystyle{mnras}

\begin{thebibliography}{}
\makeatletter
\relax
\def\mn@urlcharsother{\let\do\@makeother \do\$\do\&\do\#\do\^\do\_\do\%\do\~}
\def\mn@doi{\begingroup\mn@urlcharsother \@ifnextchar [ {\mn@doi@}
  {\mn@doi@[]}}
\def\mn@doi@[#1]#2{\def\@tempa{#1}\ifx\@tempa\@empty \href
  {http://dx.doi.org/#2} {doi:#2}\else \href {http://dx.doi.org/#2} {#1}\fi
  \endgroup}
\def\mn@eprint#1#2{\mn@eprint@#1:#2::\@nil}
\def\mn@eprint@arXiv#1{\href {http://arxiv.org/abs/#1} {{\tt arXiv:#1}}}
\def\mn@eprint@dblp#1{\href {http://dblp.uni-trier.de/rec/bibtex/#1.xml}
  {dblp:#1}}
\def\mn@eprint@#1:#2:#3:#4\@nil{\def\@tempa {#1}\def\@tempb {#2}\def\@tempc
  {#3}\ifx \@tempc \@empty \let \@tempc \@tempb \let \@tempb \@tempa \fi \ifx
  \@tempb \@empty \def\@tempb {arXiv}\fi \@ifundefined
  {mn@eprint@\@tempb}{\@tempb:\@tempc}{\expandafter \expandafter \csname
  mn@eprint@\@tempb\endcsname \expandafter{\@tempc}}}

\bibitem[\protect\citeauthoryear{{Alcolea}, {Bujarrabal}, {S{\'a}nchez
  Contreras}, {Neri}  \& {Zweigle}}{{Alcolea} et~al.}{2001}]{alcolea2001}
{Alcolea} J.,  {Bujarrabal} V.,  {S{\'a}nchez Contreras} C.,  {Neri} R.,
  {Zweigle} J.,  2001, \mn@doi [\aap] {10.1051/0004-6361:20010535}, \href
  {https://ui.adsabs.harvard.edu/abs/2001A&A...373..932A} {373, 932}

\bibitem[\protect\citeauthoryear{{Alves}, {Frau}, {Girart}, {Franco}, {Santos}
  \& {Wiesemeyer}}{{Alves} et~al.}{2014}]{alves2014}
{Alves} F.~O.,  {Frau} P.,  {Girart} J.~M.,  {Franco} G.~A.~P.,  {Santos}
  F.~P.,   {Wiesemeyer} H.,  2014, \mn@doi [\aap]
  {10.1051/0004-6361/201424678}, \href
  {https://ui.adsabs.harvard.edu/abs/2014A&A...569L...1A} {569, L1}

\bibitem[\protect\citeauthoryear{{Alves}, {Frau}, {Girart}, {Franco}, {Santos}
  \& {Wiesemeyer}}{{Alves} et~al.}{2015}]{alves2015}
{Alves} F.~O.,  {Frau} P.,  {Girart} J.~M.,  {Franco} G.~A.~P.,  {Santos}
  F.~P.,   {Wiesemeyer} H.,  2015, {On the radiation driven alignment of dust
  grains: Detection of the polarization hole in a starless core (Corrigendum)},
  Astronomy \& Astrophysics, Volume 574, id.C4, 1 pp.,
  \mn@doi{10.1051/0004-6361/201424678e}

\bibitem[\protect\citeauthoryear{{Andersson}, {Lazarian}  \&
  {Vaillancourt}}{{Andersson} et~al.}{2015}]{andersson2015}
{Andersson} B.~G.,  {Lazarian} A.,   {Vaillancourt} J.~E.,  2015, \mn@doi
  [\araa] {10.1146/annurev-astro-082214-122414}, \href
  {https://ui.adsabs.harvard.edu/abs/2015ARA&A..53..501A} {53, 501}

\bibitem[\protect\citeauthoryear{{Andersson}, {Karoly}, {Bastien}, {Soam},
  {Coud{\'e}}, {Tahani}, {Gordon}  \& {Fox-Middleton}}{{Andersson}
  et~al.}{2024}]{andersson2024}
{Andersson} B.~G.,  {Karoly} J.,  {Bastien} P.,  {Soam} A.,  {Coud{\'e}} S.,
  {Tahani} M.,  {Gordon} M.~S.,   {Fox-Middleton} S.,  2024, \mn@doi [\apj]
  {10.3847/1538-4357/ad1835}, \href
  {https://ui.adsabs.harvard.edu/abs/2024ApJ...963...76A} {963, 76}

\bibitem[\protect\citeauthoryear{{Astropy Collaboration} et~al.,}{{Astropy
  Collaboration} et~al.}{2013}]{astropy:2013}
{Astropy Collaboration} et~al., 2013, \mn@doi [\aap]
  {10.1051/0004-6361/201322068}, \href
  {http://adsabs.harvard.edu/abs/2013A%26A...558A..33A} {558, A33}

\bibitem[\protect\citeauthoryear{{Astropy Collaboration} et~al.,}{{Astropy
  Collaboration} et~al.}{2018}]{astropy:2018}
{Astropy Collaboration} et~al., 2018, \mn@doi [\aj] {10.3847/1538-3881/aabc4f},
  \href {https://ui.adsabs.harvard.edu/abs/2018AJ....156..123A} {156, 123}

\bibitem[\protect\citeauthoryear{{Astropy Collaboration} et~al.,}{{Astropy
  Collaboration} et~al.}{2022}]{astropy:2022}
{Astropy Collaboration} et~al., 2022, \mn@doi [apj] {10.3847/1538-4357/ac7c74},
  \href {https://ui.adsabs.harvard.edu/abs/2022ApJ...935..167A} {935, 167}

\bibitem[\protect\citeauthoryear{{Balick}, {Huarte-Espinosa}, {Frank}, {Gomez},
  {Alcolea}, {Corradi}  \& {Vinkovi{\'c}}}{{Balick} et~al.}{2013}]{balick2013}
{Balick} B.,  {Huarte-Espinosa} M.,  {Frank} A.,  {Gomez} T.,  {Alcolea} J.,
  {Corradi} R. L.~M.,   {Vinkovi{\'c}} D.,  2013, \mn@doi [\apj]
  {10.1088/0004-637X/772/1/20}, \href
  {https://ui.adsabs.harvard.edu/abs/2013ApJ...772...20B} {772, 20}

\bibitem[\protect\citeauthoryear{{Balick}, {Frank}, {Liu}  \&
  {Huarte-Espinosa}}{{Balick} et~al.}{2017}]{balick2017}
{Balick} B.,  {Frank} A.,  {Liu} B.,   {Huarte-Espinosa} M.,  2017, \mn@doi
  [\apj] {10.3847/1538-4357/aa77f0}, \href
  {https://ui.adsabs.harvard.edu/abs/2017ApJ...843..108B} {843, 108}

\bibitem[\protect\citeauthoryear{{Balick}, {Frank}  \& {Liu}}{{Balick}
  et~al.}{2019}]{balick2019}
{Balick} B.,  {Frank} A.,   {Liu} B.,  2019, \mn@doi [\apj]
  {10.3847/1538-4357/ab16f5}, \href
  {https://ui.adsabs.harvard.edu/abs/2019ApJ...877...30B} {877, 30}

\bibitem[\protect\citeauthoryear{{Beckwith}, {Sargent}, {Chini}  \&
  {Guesten}}{{Beckwith} et~al.}{1990}]{beckwith1990}
{Beckwith} S. V.~W.,  {Sargent} A.~I.,  {Chini} R.~S.,   {Guesten} R.,  1990,
  \mn@doi [\aj] {10.1086/115385}, \href
  {https://ui.adsabs.harvard.edu/abs/1990AJ.....99..924B} {99, 924}

\bibitem[\protect\citeauthoryear{{Bujarrabal}, {Gomez-Gonzalez}, {Bachiller}
  \& {Martin-Pintado}}{{Bujarrabal} et~al.}{1988}]{bujarrabal1988}
{Bujarrabal} V.,  {Gomez-Gonzalez} J.,  {Bachiller} R.,   {Martin-Pintado} J.,
  1988, \aap, \href {https://ui.adsabs.harvard.edu/abs/1988A&A...204..242B}
  {204, 242}

\bibitem[\protect\citeauthoryear{{Bujarrabal}, {Alcolea}, {S{\'a}nchez
  Contreras}  \& {Sahai}}{{Bujarrabal} et~al.}{2002}]{bujarrabal2002}
{Bujarrabal} V.,  {Alcolea} J.,  {S{\'a}nchez Contreras} C.,   {Sahai} R.,
  2002, \mn@doi [\aap] {10.1051/0004-6361:20020455}, \href
  {https://ui.adsabs.harvard.edu/abs/2002A&A...389..271B} {389, 271}

\bibitem[\protect\citeauthoryear{{Burgarella} et~al.,}{{Burgarella}
  et~al.}{2024}]{burgarella2024}
{Burgarella} D.,  et~al., 2024, in {Coyle} L.~E.,  {Matsuura} S.,   {Perrin}
  M.~D.,  eds,  Society of Photo-Optical Instrumentation Engineers (SPIE)
  Conference Series Vol. 13092, Space Telescopes and Instrumentation 2024:
  Optical, Infrared, and Millimeter Wave. p. 130923B,
  \mn@doi{10.1117/12.3018024}

\bibitem[\protect\citeauthoryear{{Cernicharo}, {Guelin}, {Martin-Pintado},
  {Penalver}  \& {Mauersberger}}{{Cernicharo} et~al.}{1989}]{cernicharo1989}
{Cernicharo} J.,  {Guelin} M.,  {Martin-Pintado} J.,  {Penalver} J.,
  {Mauersberger} R.,  1989, \aap, \href
  {https://ui.adsabs.harvard.edu/abs/1989A&A...222L...1C} {222, L1}

\bibitem[\protect\citeauthoryear{{Chapin}, {Berry}, {Gibb}, {Jenness}, {Scott},
  {Tilanus}, {Economou}  \& {Holland}}{{Chapin} et~al.}{2013}]{chapin2013}
{Chapin} E.~L.,  {Berry} D.~S.,  {Gibb} A.~G.,  {Jenness} T.,  {Scott} D.,
  {Tilanus} R. P.~J.,  {Economou} F.,   {Holland} W.~S.,  2013, \mn@doi
  [\mnras] {10.1093/mnras/stt052}, \href
  {https://ui.adsabs.harvard.edu/abs/2013MNRAS.430.2545C} {430, 2545}

\bibitem[\protect\citeauthoryear{{Chiar}, {Pendleton}, {Geballe}  \&
  {Tielens}}{{Chiar} et~al.}{1998}]{chiar1998}
{Chiar} J.~E.,  {Pendleton} Y.~J.,  {Geballe} T.~R.,   {Tielens} A.~G.~G.~M.,
  1998, \mn@doi [\apj] {10.1086/306318}, \href
  {https://ui.adsabs.harvard.edu/abs/1998ApJ...507..281C} {507, 281}

\bibitem[\protect\citeauthoryear{{Choi}, {Brunthaler}, {Menten}  \&
  {Reid}}{{Choi} et~al.}{2012}]{choi2012}
{Choi} Y.~K.,  {Brunthaler} A.,  {Menten} K.~M.,   {Reid} M.~J.,  2012, in
  {Booth} R.~S.,  {Vlemmings} W. H.~T.,   {Humphreys} E. M.~L.,  eds,
  {Proceedings of the International Astronomical Union} Vol. 287, {Cosmic
  Masers - from OH to H0}. pp 407--410, \mn@doi{10.1017/S1743921312007387}

\bibitem[\protect\citeauthoryear{{Cotton}}{{Cotton}}{2017}]{cotton2017}
{Cotton} W.~D.,  2017, \mn@doi [\pasp] {10.1088/1538-3873/aa793f}, \href
  {https://ui.adsabs.harvard.edu/abs/2017PASP..129i4501C} {129, 094501}

\bibitem[\protect\citeauthoryear{{Currie}, {Berry}, {Jenness}, {Gibb}, {Bell}
  \& {Draper}}{{Currie} et~al.}{2014}]{currie2014}
{Currie} M.~J.,  {Berry} D.~S.,  {Jenness} T.,  {Gibb} A.~G.,  {Bell} G.~S.,
  {Draper} P.~W.,  2014, in {Manset} N.,  {Forshay} P.,  eds,  Astronomical
  Society of the Pacific Conference Series Vol. 485, Astronomical Data Analysis
  Software and Systems XXIII. p.~391

\bibitem[\protect\citeauthoryear{{Dempsey} et~al.,}{{Dempsey}
  et~al.}{2013}]{dempsey2013}
{Dempsey} J.~T.,  et~al., 2013, \mn@doi [\mnras] {10.1093/mnras/stt090}, \href
  {https://ui.adsabs.harvard.edu/abs/2013MNRAS.430.2534D} {430, 2534}

\bibitem[\protect\citeauthoryear{{Dowell}, {Hensley}  \& {Sauvage}}{{Dowell}
  et~al.}{2024}]{dowell2024}
{Dowell} C.~D.,  {Hensley} B.~S.,   {Sauvage} M.,  2024, \mn@doi [arXiv
  e-prints] {10.48550/arXiv.2404.17050}, \href
  {https://ui.adsabs.harvard.edu/abs/2024arXiv240417050D} {p. arXiv:2404.17050}

\bibitem[\protect\citeauthoryear{{Duthu}, {Herpin}, {Wiesemeyer}, {Baudry},
  {L{\`e}bre}  \& {Paubert}}{{Duthu} et~al.}{2017}]{duthu2017}
{Duthu} A.,  {Herpin} F.,  {Wiesemeyer} H.,  {Baudry} A.,  {L{\`e}bre} A.,
  {Paubert} G.,  2017, \mn@doi [\aap] {10.1051/0004-6361/201730485}, \href
  {https://ui.adsabs.harvard.edu/abs/2017A&A...604A..12D} {604, A12}

\bibitem[\protect\citeauthoryear{{Friberg}, {Bastien}, {Berry}, {Savini},
  {Graves}  \& {Pattle}}{{Friberg} et~al.}{2016}]{friberg2016}
{Friberg} P.,  {Bastien} P.,  {Berry} D.,  {Savini} G.,  {Graves} S.~F.,
  {Pattle} K.,  2016, in {Holland} W.~S.,  {Zmuidzinas} J.,  eds,  Society of
  Photo-Optical Instrumentation Engineers (SPIE) Conference Series Vol. 9914,
  Millimeter, Submillimeter, and Far-Infrared Detectors and Instrumentation for
  Astronomy VIII. p. 991403, \mn@doi{10.1117/12.2231943}

\bibitem[\protect\citeauthoryear{{Goodman}, {Jones}, {Lada}  \&
  {Myers}}{{Goodman} et~al.}{1992}]{goodman1992}
{Goodman} A.~A.,  {Jones} T.~J.,  {Lada} E.~A.,   {Myers} P.~C.,  1992, \mn@doi
  [\apj] {10.1086/171907}, \href
  {https://ui.adsabs.harvard.edu/abs/1992ApJ...399..108G} {399, 108}

\bibitem[\protect\citeauthoryear{{Goodrich}}{{Goodrich}}{1991}]{goodrich1991}
{Goodrich} R.~W.,  1991, \mn@doi [\apj] {10.1086/170313}, \href
  {https://ui.adsabs.harvard.edu/abs/1991ApJ...376..654G} {376, 654}

\bibitem[\protect\citeauthoryear{{Greaves}}{{Greaves}}{2002}]{greaves2002}
{Greaves} J.~S.,  2002, \mn@doi [\aap] {10.1051/0004-6361:20021002}, \href
  {https://ui.adsabs.harvard.edu/abs/2002A&A...392L...1G} {392, L1}

\bibitem[\protect\citeauthoryear{{Habing} \& {Olofsson}}{{Habing} \&
  {Olofsson}}{2004}]{habing2004}
{Habing} H.~J.,  {Olofsson} H.,  2004, {Asymptotic Giant Branch Stars}.
{Springer Astronomy and Astrophysics Library},
  \mn@doi{10.1007/978-1-4757-3876-6}

\bibitem[\protect\citeauthoryear{{Hildebrand}}{{Hildebrand}}{1983}]{hildebrand1983}
{Hildebrand} R.~H.,  1983, \qjras, \href
  {https://ui.adsabs.harvard.edu/abs/1983QJRAS..24..267H} {24, 267}

\bibitem[\protect\citeauthoryear{{Hoang} \& {Lazarian}}{{Hoang} \&
  {Lazarian}}{2016}]{hoang2016}
{Hoang} T.,  {Lazarian} A.,  2016, \mn@doi [\apj]
  {10.3847/0004-637X/831/2/159}, \href
  {https://ui.adsabs.harvard.edu/abs/2016ApJ...831..159H} {831, 159}

\bibitem[\protect\citeauthoryear{{Hoang}, {Tram}, {Lee}, {Diep}  \&
  {Ngoc}}{{Hoang} et~al.}{2021}]{hoang2021}
{Hoang} T.,  {Tram} L.~N.,  {Lee} H.,  {Diep} P.~N.,   {Ngoc} N.~B.,  2021,
  \mn@doi [\apj] {10.3847/1538-4357/abd54f}, \href
  {https://ui.adsabs.harvard.edu/abs/2021ApJ...908..218H} {908, 218}

\bibitem[\protect\citeauthoryear{{Holland} et~al.,}{{Holland}
  et~al.}{2013}]{holland2013}
{Holland} W.~S.,  et~al., 2013, \mn@doi [\mnras] {10.1093/mnras/sts612}, \href
  {https://ui.adsabs.harvard.edu/abs/2013MNRAS.430.2513H} {430, 2513}

\bibitem[\protect\citeauthoryear{{Huang}, {Lee}, {Moraghan}  \&
  {Smith}}{{Huang} et~al.}{2016}]{huang2016}
{Huang} P.-S.,  {Lee} C.-F.,  {Moraghan} A.,   {Smith} M.,  2016, \mn@doi
  [\apj] {10.3847/0004-637X/820/2/134}, \href
  {https://ui.adsabs.harvard.edu/abs/2016ApJ...820..134H} {820, 134}

\bibitem[\protect\citeauthoryear{{Ishihara}, {Kaneda}, {Onaka}, {Ita},
  {Matsuura}  \& {Matsunaga}}{{Ishihara} et~al.}{2011}]{ishihara2011}
{Ishihara} D.,  {Kaneda} H.,  {Onaka} T.,  {Ita} Y.,  {Matsuura} M.,
  {Matsunaga} N.,  2011, \mn@doi [\aap] {10.1051/0004-6361/201117626}, \href
  {https://ui.adsabs.harvard.edu/abs/2011A&A...534A..79I} {534, A79}

\bibitem[\protect\citeauthoryear{{Jones}, {Bagley}, {Krejny}, {Andersson}  \&
  {Bastien}}{{Jones} et~al.}{2015}]{jones2015}
{Jones} T.~J.,  {Bagley} M.,  {Krejny} M.,  {Andersson} B.~G.,   {Bastien} P.,
  2015, \mn@doi [\aj] {10.1088/0004-6256/149/1/31}, \href
  {https://ui.adsabs.harvard.edu/abs/2015AJ....149...31J} {149, 31}

\bibitem[\protect\citeauthoryear{{Karoly} et~al.,}{{Karoly}
  et~al.}{2023}]{karoly2023}
{Karoly} J.,  et~al., 2023, \mn@doi [\apj] {10.3847/1538-4357/acd6f2}, \href
  {https://ui.adsabs.harvard.edu/abs/2023ApJ...952...29K} {952, 29}

\bibitem[\protect\citeauthoryear{{Klaassen} et~al.,}{{Klaassen}
  et~al.}{2024}]{klaassen2024}
{Klaassen} P.,  et~al., 2024, \mn@doi [Open Research Europe]
  {10.12688/openreseurope.17450.1}, \href
  {https://ui.adsabs.harvard.edu/abs/2024ORE.....4..112K} {4, 112}

\bibitem[\protect\citeauthoryear{{Knapp}, {Sandell}  \& {Robson}}{{Knapp}
  et~al.}{1993}]{knapp1993}
{Knapp} G.~R.,  {Sandell} G.,   {Robson} E.~I.,  1993, \mn@doi [\apjs]
  {10.1086/191820}, \href
  {https://ui.adsabs.harvard.edu/abs/1993ApJS...88..173K} {88, 173}

\bibitem[\protect\citeauthoryear{{Kwok}}{{Kwok}}{1993}]{kwok1993}
{Kwok} S.,  1993, \mn@doi [\araa] {10.1146/annurev.aa.31.090193.000431}, \href
  {https://ui.adsabs.harvard.edu/abs/1993ARA&A..31...63K} {31, 63}

\bibitem[\protect\citeauthoryear{{Kwok} \& {Bignell}}{{Kwok} \&
  {Bignell}}{1984}]{kwok1984}
{Kwok} S.,  {Bignell} R.~C.,  1984, \mn@doi [\apj] {10.1086/161643}, \href
  {https://ui.adsabs.harvard.edu/abs/1984ApJ...276..544K} {276, 544}

\bibitem[\protect\citeauthoryear{{Lazarian} \& {Hoang}}{{Lazarian} \&
  {Hoang}}{2007}]{lazarian2007}
{Lazarian} A.,  {Hoang} T.,  2007, \mn@doi [\mnras]
  {10.1111/j.1365-2966.2007.11817.x}, \href
  {https://ui.adsabs.harvard.edu/abs/2007MNRAS.378..910L} {378, 910}

\bibitem[\protect\citeauthoryear{{Leal-Ferreira}, {Vlemmings}, {Diamond},
  {Kemball}, {Amiri}  \& {Desmurs}}{{Leal-Ferreira}
  et~al.}{2012}]{lealferreira2012}
{Leal-Ferreira} M.~L.,  {Vlemmings} W.~H.~T.,  {Diamond} P.~J.,  {Kemball} A.,
  {Amiri} N.,   {Desmurs} J.~F.,  2012, \mn@doi [\aap]
  {10.1051/0004-6361/201118303}, \href
  {https://ui.adsabs.harvard.edu/abs/2012A&A...540A..42L} {540, A42}

\bibitem[\protect\citeauthoryear{{Lee}, {Sahai}, {S{\'a}nchez Contreras},
  {Huang}  \& {Hao Tay}}{{Lee} et~al.}{2013}]{lee2013}
{Lee} C.-F.,  {Sahai} R.,  {S{\'a}nchez Contreras} C.,  {Huang} P.-S.,   {Hao
  Tay} J.~J.,  2013, \mn@doi [\apj] {10.1088/0004-637X/777/1/37}, \href
  {https://ui.adsabs.harvard.edu/abs/2013ApJ...777...37L} {777, 37}

\bibitem[\protect\citeauthoryear{{Lyo} et~al.,}{{Lyo} et~al.}{2021}]{lyo2021}
{Lyo} A.~R.,  et~al., 2021, \mn@doi [\apj] {10.3847/1538-4357/ac0ce9}, \href
  {https://ui.adsabs.harvard.edu/abs/2021ApJ...918...85L} {918, 85}

\bibitem[\protect\citeauthoryear{{Mairs} et~al.,}{{Mairs}
  et~al.}{2021}]{mairs2021}
{Mairs} S.,  et~al., 2021, \mn@doi [\aj] {10.3847/1538-3881/ac18bf}, \href
  {https://ui.adsabs.harvard.edu/abs/2021AJ....162..191M} {162, 191}

\bibitem[\protect\citeauthoryear{{Montier}, {Plaszczynski}, {Levrier},
  {Tristram}, {Alina}, {Ristorcelli}, {Bernard}  \& {Guillet}}{{Montier}
  et~al.}{2015}]{montier2015}
{Montier} L.,  {Plaszczynski} S.,  {Levrier} F.,  {Tristram} M.,  {Alina} D.,
  {Ristorcelli} I.,  {Bernard} J.~P.,   {Guillet} V.,  2015, \mn@doi [\aap]
  {10.1051/0004-6361/201424451}, \href
  {https://ui.adsabs.harvard.edu/abs/2015A&A...574A.136M} {574, A136}

\bibitem[\protect\citeauthoryear{{Morris}, {Guilloteau}, {Lucas}  \&
  {Omont}}{{Morris} et~al.}{1987}]{morris1987}
{Morris} M.,  {Guilloteau} S.,  {Lucas} R.,   {Omont} A.,  1987, \mn@doi [\apj]
  {10.1086/165681}, \href
  {https://ui.adsabs.harvard.edu/abs/1987ApJ...321..888M} {321, 888}

\bibitem[\protect\citeauthoryear{{Mroczkowski} et~al.,}{{Mroczkowski}
  et~al.}{2025}]{mroczkowski2025}
{Mroczkowski} T.,  et~al., 2025, \mn@doi [\aap] {10.1051/0004-6361/202449786},
  \href {https://ui.adsabs.harvard.edu/abs/2025A&A...694A.142M} {694, A142}

\bibitem[\protect\citeauthoryear{{Murakawa}, {Izumiura}, {Oudmaijer}  \&
  {Maud}}{{Murakawa} et~al.}{2013}]{murakawa2013}
{Murakawa} K.,  {Izumiura} H.,  {Oudmaijer} R.~D.,   {Maud} L.~T.,  2013,
  \mn@doi [\mnras] {10.1093/mnras/stt118}, \href
  {https://ui.adsabs.harvard.edu/abs/2013MNRAS.430.3112M} {430, 3112}

\bibitem[\protect\citeauthoryear{{Naghizadeh-Khouei} \&
  {Clarke}}{{Naghizadeh-Khouei} \& {Clarke}}{1993}]{naghizadehkhouei1993}
{Naghizadeh-Khouei} J.,  {Clarke} D.,  1993, \aap, \href
  {https://ui.adsabs.harvard.edu/abs/1993A&A...274..968N} {274, 968}

\bibitem[\protect\citeauthoryear{{Nishikawa}, {Takami}, {Hayashi}, {Wiseman}
  \& {Pyo}}{{Nishikawa} et~al.}{2008}]{nishikawa2008}
{Nishikawa} T.,  {Takami} M.,  {Hayashi} M.,  {Wiseman} J.,   {Pyo} T.-S.,
  2008, \mn@doi [\apj] {10.1086/588644}, \href
  {https://ui.adsabs.harvard.edu/abs/2008ApJ...684.1260N} {684, 1260}

\bibitem[\protect\citeauthoryear{{Parthasarathy} \& {Pottasch}}{{Parthasarathy}
  \& {Pottasch}}{1986}]{parthasarathy1986}
{Parthasarathy} M.,  {Pottasch} S.~R.,  1986, \aap, \href
  {https://ui.adsabs.harvard.edu/abs/1986A&A...154L..16P} {154, L16}

\bibitem[\protect\citeauthoryear{{Pattle} et~al.,}{{Pattle}
  et~al.}{2019}]{pattle2019a}
{Pattle} K.,  et~al., 2019, \mn@doi [\apj] {10.3847/1538-4357/ab286f}, \href
  {https://ui.adsabs.harvard.edu/abs/2019ApJ...880...27P} {880, 27}

\bibitem[\protect\citeauthoryear{{Pattle} et~al.,}{{Pattle}
  et~al.}{2021}]{pattle2021}
{Pattle} K.,  et~al., 2021, \mn@doi [\apj] {10.3847/1538-4357/abcc6c}, \href
  {https://ui.adsabs.harvard.edu/abs/2021ApJ...907...88P} {907, 88}

\bibitem[\protect\citeauthoryear{{Pattle} et~al.,}{{Pattle}
  et~al.}{2022}]{pattle2022}
{Pattle} K.,  et~al., 2022, \mn@doi [\mnras] {10.1093/mnras/stac1356}, \href
  {https://ui.adsabs.harvard.edu/abs/2022MNRAS.515.1026P} {515, 1026}

\bibitem[\protect\citeauthoryear{{Planck Collaboration} et~al.,}{{Planck
  Collaboration} et~al.}{2014}]{planck2014xi}
{Planck Collaboration} et~al., 2014, \mn@doi [\aap]
  {10.1051/0004-6361/201323195}, \href
  {https://ui.adsabs.harvard.edu/abs/2014A&A...571A..11P} {571, A11}

\bibitem[\protect\citeauthoryear{{Plaszczynski}, {Montier}, {Levrier}  \&
  {Tristram}}{{Plaszczynski} et~al.}{2014}]{plaszczynski2014}
{Plaszczynski} S.,  {Montier} L.,  {Levrier} F.,   {Tristram} M.,  2014,
  \mn@doi [\mnras] {10.1093/mnras/stu270}, \href
  {https://ui.adsabs.harvard.edu/abs/2014MNRAS.439.4048P} {439, 4048}

\bibitem[\protect\citeauthoryear{Privon, Cerrigone, Corvillon, Kawamura,
  Popping  \& Rebolledo}{Privon et~al.}{2025}]{privon2025}
Privon G.,  Cerrigone L.,  Corvillon A.,  Kawamura A.,  Popping G.,   Rebolledo
  D.,  2025, ALMA Proposer’s Guide, ALMA Doc. 12.2, ver. 1.0,
  \mn@doi{10.5281/zenodo.4511961}

\bibitem[\protect\citeauthoryear{{Reipurth}}{{Reipurth}}{1987}]{reipurth1987}
{Reipurth} B.,  1987, \mn@doi [\nat] {10.1038/325787a0}, \href
  {https://ui.adsabs.harvard.edu/abs/1987Natur.325..787R} {325, 787}

\bibitem[\protect\citeauthoryear{{Sabin}, {Zhang}, {Zijlstra}, {Patel},
  {V{\'a}zquez}, {Zauderer}, {Contreras}  \& {Guill{\'e}n}}{{Sabin}
  et~al.}{2014}]{sabin2014}
{Sabin} L.,  {Zhang} Q.,  {Zijlstra} A.~A.,  {Patel} N.~A.,  {V{\'a}zquez} R.,
  {Zauderer} B.~A.,  {Contreras} M.~E.,   {Guill{\'e}n} P.~F.,  2014, \mn@doi
  [\mnras] {10.1093/mnras/stt2318}, \href
  {https://ui.adsabs.harvard.edu/abs/2014MNRAS.438.1794S} {438, 1794}

\bibitem[\protect\citeauthoryear{{Sabin}, {Hull}, {Plambeck}, {Zijlstra},
  {V{\'a}zquez}, {Navarro}  \& {Guill{\'e}n}}{{Sabin} et~al.}{2015}]{sabin2015}
{Sabin} L.,  {Hull} C.~L.~H.,  {Plambeck} R.~L.,  {Zijlstra} A.~A.,
  {V{\'a}zquez} R.,  {Navarro} S.~G.,   {Guill{\'e}n} P.~F.,  2015, \mn@doi
  [\mnras] {10.1093/mnras/stv461}, \href
  {https://ui.adsabs.harvard.edu/abs/2015MNRAS.449.2368S} {449, 2368}

\bibitem[\protect\citeauthoryear{{Sabin} et~al.,}{{Sabin}
  et~al.}{2020}]{sabin2020}
{Sabin} L.,  et~al., 2020, \mn@doi [\mnras] {10.1093/mnras/staa1449}, \href
  {https://ui.adsabs.harvard.edu/abs/2020MNRAS.495.4297S} {495, 4297}

\bibitem[\protect\citeauthoryear{{Sabin}, {Ramirez}, {Luna}, {Serrano},
  {Toal\'a}, {Garcia}  \& {Meixner}}{{Sabin} et~al.}{2023}]{sabin2023}
{Sabin} L.,  {Ramirez} E.,  {Luna} A.,  {Serrano} O.,  {Toal\'a} J.,  {Garcia}
  A.,   {Meixner} M.,  2023, in {{Moullet}, A. and {Kataria}, T. and {Lis}, D.
  and {Unwin}, S. and {Hasegawa}, Y. and {Mills}, E. and {Battersby}, C. and
  {Roc}, A. and {Meixner}, M.} ed., , {PRIMA General Observer Science Book}.
Goddard/JPL/Caltech, Chapt.~70, pp 358--364 (\mn@eprint {arXiv} {2310.20572}),
  \url {https://arxiv.org/abs/2310.20572}

\bibitem[\protect\citeauthoryear{{Sahai}, {Morris}, {S{\'a}nchez Contreras}  \&
  {Claussen}}{{Sahai} et~al.}{2007}]{sahai2007}
{Sahai} R.,  {Morris} M.,  {S{\'a}nchez Contreras} C.,   {Claussen} M.,  2007,
  \mn@doi [\aj] {10.1086/522944}, \href
  {https://ui.adsabs.harvard.edu/abs/2007AJ....134.2200S} {134, 2200}

\bibitem[\protect\citeauthoryear{{S{\'a}nchez Contreras}, {Bujarrabal},
  {Miranda}  \& {Fern{\'a}ndez-Figueroa}}{{S{\'a}nchez Contreras}
  et~al.}{2000}]{sanchezcontreras2000}
{S{\'a}nchez Contreras} C.,  {Bujarrabal} V.,  {Miranda} L.~F.,
  {Fern{\'a}ndez-Figueroa} M.~J.,  2000, \aap, \href
  {https://ui.adsabs.harvard.edu/abs/2000A&A...355.1103S} {355, 1103}

\bibitem[\protect\citeauthoryear{{S{\'a}nchez Contreras} et~al.,}{{S{\'a}nchez
  Contreras} et~al.}{2015}]{sanchezcontreras2015}
{S{\'a}nchez Contreras} C.,  et~al., 2015, \mn@doi [\aap]
  {10.1051/0004-6361/201525652}, \href
  {https://ui.adsabs.harvard.edu/abs/2015A&A...577A..52S} {577, A52}

\bibitem[\protect\citeauthoryear{{Sanchez Contreras}, {Alcolea}, {Rodriguez
  Cardoso}, {Bujarrabal}, {Castro-Carrizo}, {Quintana-Lacaci}, {Velilla-Prieto}
   \& {Santander-Garcia}}{{Sanchez Contreras}
  et~al.}{2022}]{sanchezcontreras2022}
{Sanchez Contreras} C.,  {Alcolea} J.,  {Rodriguez Cardoso} R.,  {Bujarrabal}
  V.,  {Castro-Carrizo} A.,  {Quintana-Lacaci} G.,  {Velilla-Prieto} L.,
  {Santander-Garcia} M.,  2022, \mn@doi [\aap] {10.1051/0004-6361/202243623},
  \href {https://ui.adsabs.harvard.edu/abs/2022A&A...665A..88S} {665, A88}

\bibitem[\protect\citeauthoryear{{Serkowski}}{{Serkowski}}{1958}]{serkowski1958}
{Serkowski} K.,  1958, \actaa, \href
  {https://ui.adsabs.harvard.edu/abs/1958AcA.....8..135S} {8, 135}

\bibitem[\protect\citeauthoryear{{Serkowski}}{{Serkowski}}{1962}]{serkowski1962}
{Serkowski} K.,  1962, \mn@doi [Advances in Astronomy and Astrophysics]
  {10.1016/B978-1-4831-9919-1.50009-1}, \href
  {https://ui.adsabs.harvard.edu/abs/1962AdA&A...1..289S} {1, 289}

\bibitem[\protect\citeauthoryear{{Sharma}, {Pattle}, {Li}, {Lee}, {Gopinathan},
  {Ching}, {Tahani}  \& {Kim}}{{Sharma} et~al.}{2025}]{sharma2025}
{Sharma} E.,  {Pattle} K.,  {Li} D.,  {Lee} C.~W.,  {Gopinathan} M.,  {Ching}
  T.-C.,  {Tahani} M.,   {Kim} S.,  2025, \mn@doi [arXiv e-prints]
  {10.48550/arXiv.2503.20721}, \href
  {https://ui.adsabs.harvard.edu/abs/2025arXiv250320721S} {p. arXiv:2503.20721}

\bibitem[\protect\citeauthoryear{{Simmons} \& {Stewart}}{{Simmons} \&
  {Stewart}}{1985}]{simmons1985}
{Simmons} J.~F.~L.,  {Stewart} B.~G.,  1985, \aap, \href
  {https://ui.adsabs.harvard.edu/abs/1985A&A...142..100S} {142, 100}

\bibitem[\protect\citeauthoryear{{Stanimirovic}}{{Stanimirovic}}{2002}]{stanimirovic2002}
{Stanimirovic} S.,  2002, in {Stanimirovic} S.,  {Altschuler} D.,  {Goldsmith}
  P.,   {Salter} C.,  eds,  Astronomical Society of the Pacific Conference
  Series Vol. 278, Single-Dish Radio Astronomy: Techniques and Applications. pp
  375--396 (\mn@eprint {arXiv} {astro-ph/0205329}),
  \mn@doi{10.48550/arXiv.astro-ph/0205329}

\bibitem[\protect\citeauthoryear{{Ueta}, {Meixner}  \& {Bobrowsky}}{{Ueta}
  et~al.}{2000}]{ueta2000}
{Ueta} T.,  {Meixner} M.,   {Bobrowsky} M.,  2000, \mn@doi [\apj]
  {10.1086/308208}, \href
  {https://ui.adsabs.harvard.edu/abs/2000ApJ...528..861U} {528, 861}

\bibitem[\protect\citeauthoryear{{Ward-Thompson}, {Kirk}, {Crutcher},
  {Greaves}, {Holland}  \& {Andr{\'e}}}{{Ward-Thompson}
  et~al.}{2000}]{wardthompson2000}
{Ward-Thompson} D.,  {Kirk} J.~M.,  {Crutcher} R.~M.,  {Greaves} J.~S.,
  {Holland} W.~S.,   {Andr{\'e}} P.,  2000, \mn@doi [\apjl] {10.1086/312764},
  \href {https://ui.adsabs.harvard.edu/abs/2000ApJ...537L.135W} {537, L135}

\bibitem[\protect\citeauthoryear{{Wardle} \& {Kronberg}}{{Wardle} \&
  {Kronberg}}{1974}]{wardle1974}
{Wardle} J.~F.~C.,  {Kronberg} P.~P.,  1974, \mn@doi [\apj] {10.1086/153240},
  \href {https://ui.adsabs.harvard.edu/abs/1974ApJ...194..249W} {194, 249}

\bibitem[\protect\citeauthoryear{{Whittet}, {Hough}, {Lazarian}  \&
  {Hoang}}{{Whittet} et~al.}{2008}]{whittet2008}
{Whittet} D.~C.~B.,  {Hough} J.~H.,  {Lazarian} A.,   {Hoang} T.,  2008,
  \mn@doi [\apj] {10.1086/525040}, \href
  {https://ui.adsabs.harvard.edu/abs/2008ApJ...674..304W} {674, 304}

\bibitem[\protect\citeauthoryear{{Yen} et~al.,}{{Yen} et~al.}{2021}]{yen2021}
{Yen} H.-W.,  et~al., 2021, \mn@doi [\apj] {10.3847/1538-4357/abca99}, \href
  {https://ui.adsabs.harvard.edu/abs/2021ApJ...907...33Y} {907, 33}

\makeatother
\end{thebibliography}

\bsp	
\label{lastpage}
\end{document}